\documentclass[12pt,twoside]{article}   
\usepackage[super,sort,comma]{natbib}

\usepackage{fancyhdr}		
\usepackage{graphicx}
\usepackage{multirow}
\usepackage{enumitem}
\usepackage{array}
\usepackage{color}
\usepackage{ulem}
\usepackage{comment}
\newcolumntype{L}[1]{>{\raggedright\let\newline\\\arraybackslash\hspace{0pt}}m{#1}}
\newcolumntype{C}[1]{>{\centering\let\newline\\\arraybackslash\hspace{0pt}}m{#1}}
\newcolumntype{R}[1]{>{\raggedleft\let\newline\\\arraybackslash\hspace{0pt}}m{#1}}
\usepackage{subcaption}

\usepackage[section]{placeins}   %

\usepackage{graphicx}

\makeatletter \renewcommand\@biblabel[1]{$^{#1}$} \makeatother
\newcommand{\cen}[1]{\begin{center} #1 \end{center}}

\renewcommand{\refname}{}       %

\usepackage[mathlines]{lineno}

\nolinenumbers

\usepackage{hyperref}
\hypersetup{ colorlinks,
    citecolor=blue,
    filecolor=blue,
    linkcolor=blue,
    urlcolor=blue
}

\usepackage{xcolor}

\definecolor{gray}{rgb}{0.6,0.6,0.6}
\definecolor{red}{rgb}{0.85,0,0}
\definecolor{green}{rgb}{0,0.85,0}
\definecolor{blue}{rgb}{0,0,0.85}
\definecolor{beige}{rgb}{0.92,0.87,0.78}
\usepackage[all]{hypcap}    

\begin{document}

\cen{\sf {\Large {\bfseries Bridging the gap between prostate radiology and pathology through machine learning} \\ 
\vspace*{10mm}
Indrani Bhattacharya$^{1,2,\star}$, David S. Lim$^{3,\star}$, Han Lin Aung$^{4}$ \footnote{Currently at Facebook}, Xingchen Liu$^{4}$ \footnote{Currently at Momenta.ai}, Arun Seetharaman$^{5}$ \footnote{Currently at HearVista.AI}, Christian A. Kunder$^{6}$, Wei Shao$^{1}$, Simon J. C. Soerensen$^{2,7}$, Richard E. Fan$^{2}$, Pejman Ghanouni$^{1,2}$, Katherine J. {To’o}$^{1,8}$, James D. Brooks$^{2}$, Geoffrey A. Sonn$^{1,2,\&}$, Mirabela Rusu$^{1,\&}$} \\
\vspace*{10mm}

$^{1}$Department of Radiology, Stanford University School of Medicine, Stanford, CA 94305

$^{2}$Department of Urology, Stanford University School of Medicine, Stanford, CA 94305

$^{3}$Department of Computer Science, Stanford University, Stanford, CA 94305

$^{4}$Department of Biomedical Data Science, Stanford University School of Medicine, Stanford, CA 94305

$^{5}$Department of Electrical Engineering, Stanford University, Stanford, CA 94305

$^{6}$Department of Pathology, Stanford University School of Medicine, Stanford, CA 94305

$^{7}$Department of Epidemiology and Population Health, Stanford University School of Medicine, Stanford, CA 94305

$^{8}$Department of Radiology, VA Palo Alto Health Care System, Palo Alto, CA 94304

$^{\star}$ Equal contribution as first authors

$^{\&}$ Equal contribution as senior authors

} 

\pagenumbering{roman}
\setcounter{page}{1}
\pagestyle{plain}
Authors to whom correspondence should be addressed: 
Indrani Bhattacharya (ibhatt@stanford.edu), 
Mirabela Rusu (mirabela.rusu@stanford.edu)

\begin{abstract}
{\bf Background:} Prostate cancer remains the second deadliest cancer for American men despite clinical advancements. While Magnetic Resonance Imaging (MRI) is increasingly used to guide targeted biopsies for prostate cancer diagnosis, its utility remains limited due to high rates of false positives and false negatives as well as low inter-reader agreements.

{\bf Purpose:} Machine learning methods to detect and localize cancer on prostate MRI can help standardize radiologist interpretations. However, existing machine learning methods vary not only in model architecture, but also in the ground truth labeling strategies used for model training. We compare different labeling strategies and the effects they have on the performance of different machine learning models for prostate cancer detection.

{\bf Methods:} Four different deep learning models (SPCNet, U-Net, branched U-Net, and DeepLabv3+) were trained using 75 patients with radical prostatectomy, and evaluated using 40 patients with radical prostatectomy and 275 patients with targeted biopsy. Each deep learning model was trained with four different label types: pathology-confirmed radiologist labels, pathologist labels on whole-mount histopathology images, and lesion-level and pixel-level digital pathologist labels (previously validated deep learning algorithm on histopathology images to predict pixel-level Gleason patterns) on whole-mount histopathology images. The pathologist and digital pathologist labels (collectively referred to as pathology labels) were mapped onto pre-operative MRI using an automated MRI-histopathology registration platform. 

{\bf Results:} Radiologist labels missed cancers (ROC-AUC: 0.75 - 0.84), had lower lesion volumes (\textasciitilde 75\% of pathology lesions), and lower Dice overlaps (0.24 - 0.28) when compared with pathology labels. Consequently, machine learning models trained with radiologist labels also showed inferior performance compared to models trained with pathology labels. Digital pathologist labels showed high concordance with pathologist labels of cancer (lesion ROC-AUC: 0.97 - 1, lesion Dice: 0.75 - 0.93). Machine learning models trained with digital pathologist labels had the highest lesion detection rates in the radical prostatectomy cohort (aggressive lesion ROC-AUC:  0.91 - 0.94), and had generalizable and comparable performance to pathologist label trained-models in the targeted biopsy cohort (aggressive lesion ROC-AUC: 0.87 - 0.88), irrespective of the deep learning architecture. Moreover, machine learning models trained with pixel-level digital pathologist labels were able to selectively identify aggressive and indolent cancer components in mixed lesions, which is not possible with any human-annotated label type.

{\bf Conclusions:} Machine learning models for prostate MRI interpretation that are trained with digital pathologist labels showed higher or comparable performance with pathologist label-trained models in both radical prostatectomy and targeted biopsy cohort. Digital pathologist labels can reduce challenges associated with human annotations, including labor, time, inter- and intra-reader variability, and can help bridge the gap between prostate radiology and pathology by enabling the training of reliable machine learning models to detect and localize prostate cancer on MRI.\\
\end{abstract}


\setlength{\baselineskip}{0.7cm}      

\pagenumbering{arabic}
\setcounter{page}{1}
\pagestyle{fancy}
\section{Introduction}

One in eight American men will be diagnosed in their lifetime with prostate cancer as per estimates from the American Cancer Society \citep{siegel2021cancer}. Inspite of clinical advancements, prostate cancer remains the second deadliest cancer among men in the United States \citep{siegel2021cancer}. Magnetic Resonance Imaging (MRI) is increasingly used to detect and localize prostate cancer, to guide targeted biopsies and in treatment planning \citep{liu2018adoption}. Despite the potential of MRI in detecting prostate cancer, subtle differences between benign and cancerous tissue on MRI lead to false negatives \citep{ahmed2017diagnostic, johnson2019detection}, false positives \cite{ahmed2017diagnostic} and high inter-reader variability \citep{sonn2019prostate, barentsz2016synopsis, westphalen2020variability} among radiologists. Radiologist-assigned PI-RADS (Prostate Imaging-Reporting and Data System) scores also suffer from wide variability, leading to missing or over-calling aggressive cancers \citep{stolk2019false}. Urologists and radiologists often recommend biopsy despite relatively low suspicion for cancer due to concerns for missed aggressive cancers. Moreover, MRI-guided targeted biopsies are often supplemented with systematic biopsies, increasing morbidity (infection, bleeding, pain), as well as resulting in over-treatment of indolent cancers. Selective identification of aggressive and indolent cancer on MRI could potentially help detect men with aggressive prostate cancer, and reduce unnecessary biopsies in men without cancer or with indolent prostate cancer. 

In order to standardize radiologist interpretations of prostate MRI, several machine learning methods have been developed to detect cancer, localize cancer, and characterize cancer aggressiveness using prostate MR images. Prior machine learning methods for prostate cancer detection include traditional machine learning \citep{viswanath2012central, litjens2014computer, viswanath2019comparing, mcgarry2019gleason} as well as deep learning models using MRI  \citep{sumathipala2018prostate, cao2019joint,sanyal2020automated, bhattacharya2020corrsignet,seetharaman2021automated, saha2021end}. The prior studies for automated prostate cancer detection and localization on MRI not only differ in the models used, but also in the ground truth labels used to train their models  (Table \ref{tab:prior_work}). 

The variety of labels used to train existing machine learning methods of prostate cancer detection using MRI include:
\begin{enumerate} \itemsep -5pt
    \item Radiologist outlines of PI-RADS 3 or above lesions, without pathology confirmation \citep{saha2021end, yu2020deep,hosseinzadeh2021deep};
    \item Radiologist outlines with pathology confirmation from targeted biopsy \citep{sanyal2020automated};
    \item Radiologist outlines with pathology confirmation from post-operative whole-mount histopathology images of radical prostatectomy patients through cognitive registration or manual matching \citep{sumathipala2018prostate, cao2019joint};
    \item Pathologist outlines on whole-mount histopathology images mapped onto pre-operative MRI through semi-automatic or manual registration \citep{ mcgarry2019gleason};
    \item Pathologist outlines on whole-mount histopathology images mapped onto pre-operative MRI using automated MRI-histopathology registration \citep{bhattacharya2020corrsignet};
    \item Gleason pattern labels on whole-mount histopathology images derived from a previously validated deep learning algorithm \citep{ryu2019automated} mapped onto MRI through automated MRI-histopathology registration \citep{seetharaman2021automated, bhattacharya2021selective};
    
\end{enumerate}

Although different label types have been used in prior studies, no prior study investigated the comparative performance of the different label types to ascertain which labels provide the optimum training to machine learning methods applied to prostate MR images. All the label types used in prior studies have advantages as well as disadvantages. First, radiologist outlines without pathology confirmation are easier to obtain in large numbers from routine clinical care, but they include many false positives and may also miss cancers. Prior studies have shown that the false positive rate of radiologist outlines with PI-RADS scores $\geq$ 3 can vary from 32\% to 50\% \citep{stolk2019false}, depending on the experience of the radiologist. Moreover, radiologists can miss up to 12\% of aggressive cancers during screening and 34\% of aggressive cancers in men undergoing radical prostatectomy \citep{ahmed2017diagnostic, johnson2019detection}. Second, radiologist outlines with pathology confirmation (through targeted biopsy) may still miss MRI-invisible or hardly-visible lesions and underestimate tumor extent \citep{priester2017magnetic}. Third, cognitive registration or manual matching with post-operative whole-mount histopathology images of radical prostatectomy patients provides more accurate pixel-level cancer-mapping from histopathology images to pre-operative MRI, but the cancer extent is still under-estimated \citep{priester2017magnetic}, and it is still challenging to outline the \textasciitilde 20\% of tumors that are hardly-visible or invisible on MRI \citep{barentsz2016synopsis}. Fourth, pathologist labels mapped through registration onto MRI are the most accurate, but manual and semi-automatic registration are labor-intensive, time-consuming and require highly-skilled experts in both radiology and pathology \citep{kalavagunta2015registration, hurrell2017optimized, losnegaard2018intensity}. Fifth, pathologist labels mapped onto MRI using automated MRI-histopathology \citep{rusu2020registration, shao2021prosregnet, shao2021weakly, sood20213d} registration can alleviate the challenges associated with manual or semi-automatic registration approaches, but it is still challenging for human pathologists to annotate large datasets of whole-mount histopathology images with pixel-level annotations of cancer and Gleason patterns to train machine learning models on prostate MRI. Also, there can be variability in inter- and intra- pathologist assignment of Gleason grade groups. 

In this study, we compare the different labeling strategies and analyze their effects in training machine learning methods for prostate cancer detection on MRI. Since a variety of machine learning model architectures have been used in existing studies, for simplicity of discussion, in this study, we use the general term ``digital radiologists'' to refer to all deep learning models that are applied to prostate MR images to detect and localize cancer. Similarly, for simplicity, we use the term ``digital pathologists'' to refer to all deep learning models applied to prostate histopathology images for detecting cancer and assigning Gleason patterns. We use the term ``pathology labels'' to collectively refer to labels on whole-mount prostate histopathology images, derived either through human or digital pathologist annotations. To better understand the optimum approach for training reliable machine learning methods for prostate cancer, in this study, we seek answers to the following questions: (1) What effect does each label type have on the digital radiologist model they train?  (2) What is the best way to train digital radiologist models? (3) Can digital pathologists be used to train reliable digital radiologists? 

We hypothesize that digital pathologist annotations with pixel-level histologic grade labels mapped onto MRI through automated MRI-histopathology registration can (a) alleviate challenges associated with radiologist and pathologist labels, and (b) provide the most reliable digital radiologists for selective identification of aggressive and indolent prostate cancers. Recent studies have shown that digital pathologists have very high accuracy in Gleason grading on prostate histopathology images, and can significantly improve Gleason grading by pathologists by reducing variability in inter- and intra-pathologist Gleason grade group assignment \citep{ryu2019automated, bulten2020artificial, bulten2020automated}. Our prior SPCNet\citep{seetharaman2021automated} and CorrSigNIA \citep{bhattacharya2021selective} studies  are the only studies that used digital pathologist labels for training digital radiologists. 

In order to study the effects of different labeling strategies on digital radiologists,
we trained four different deep learning networks (SPCNet \cite{seetharaman2021automated}, U-Net \cite{ronneberger2015u, sanyal2020automated}, branched U-Net \citep{bhattacharya2021selective}, and DeepLabv3+ \cite{cao2019joint}) commonly used for prostate cancer detection and localization in prior studies. 
For each network architecture, we trained four different digital radiologist models using radical prostatectomy patients with four different types of labels: pathology-confirmed radiologist labels ($\mathcal{L}^{Rad}$), pathologist labels mapped to MRI through automated registration ($\mathcal{L}^{Path}$), and two variants of digital pathologist labels mapped to MRI using automated registration, lesion-level digital pathologist labels ($\mathcal{L}^{DPath}_{Lesion}$) and pixel-level digital pathologist labels ($\mathcal{L}^{DPath}_{Pixel}$). Each label type selectively identified aggressive and indolent cancer on either a lesion-level ($\mathcal{L}^{Rad}$, $\mathcal{L}^{Path}$, $\mathcal{L}^{DPath}_{Lesion}$) or a pixel-level ($\mathcal{L}^{DPath}_{Pixel}$). Selective identification on a lesion-level enables identifying entire lesions as aggressive or indolent, whereas selective identification on a pixel-level enables identifying and localizing aggressive and indolent cancer components in mixed lesions. We evaluated our trained digital radiologists in two different patient cohorts (N = 315), including 40 men with radical prostatectomy and 275 men with targeted biopsies. Evaluation on two different cohorts enabled (1) comparing the effect of different labeling strategies on digital radiologist performance, and (2) testing the generalizability of the different models. Moreover, to ascertain if the effect of the labels is independent of the model type used, we used four different deep learning algorithms to train our digital radiologists (SPCNet \cite{seetharaman2021automated}, U-Net \cite{ronneberger2015u, sanyal2020automated}, branched U-Net, and DeepLabv3+ \cite{cao2019joint}).

To summarize, the novel contributions of our study are:
\begin{enumerate} \itemsep -5pt
    \item We analysed different labeling strategies to identify the best way to train digital radiologists for selective identification of aggressive and indolent prostate cancer using MRI. 
    \item We assessed performance of digital pathologist labels and of the digital radiologists trained with these labels in comparison with human radiologist and pathologist labels. 
    \item We study whether the effect of different labeling strategies is independent of the model architecture.
    \item We study whether the effect of different labeling strategies is consistent across different patient populations with different distributions of cancer.
    
\end{enumerate}
\begin{table}[!htb]
\centering
\caption{Summary of prior machine learning methods for prostate cancer detection and localization on MRI. Abbreviations used: PCa: Prostate Cancer; RP: Radical Prostatectomy; MRI: Magnetic Resonance Imaging; DL: Deep Learning; TML: Traditional Machine Learning; FPN: Feature Pyramid Network; SPCNet: Stanford Prostate Cancer Network.}
\label{tab:prior_work}
\resizebox{\textwidth}{!}{%
\begin{tabular}{|l|l|l|l|l|l|}
\hline
\textbf{Prior study} & \textbf{Method} &
  \textbf{Label type} &
  \textbf{\begin{tabular}[c]{@{}l@{}}Pathology\\ confirmation\end{tabular}} &
  \textbf{\begin{tabular}[c]{@{}l@{}}Pathology \\ type\end{tabular}} &
  \textbf{\begin{tabular}[c]{@{}l@{}}Mapping from pathology\\  to MRI, if applicable\end{tabular}} \\ \hline
Saha et al.~\citep{saha2021end} &  \begin{tabular}[c]{@{}l@{}}DL (U-Net variant \\+ residual classifier)\end{tabular}
 &
  Radiologist &
  No &
  N/A & N/A
   \\ \hline
   Yu et al.~\citep{yu2020deep} & \begin{tabular}[c]{@{}l@{}}DL (ResNet + Panoptic FPN \\+ Mask R-CNN\\ + Attention module)\end{tabular} &
  Radiologist &
  No &
  N/A & N/A
   \\ \hline
   Hosseinzadeh et al.~\citep{hosseinzadeh2021deep} &  \begin{tabular}[c]{@{}l@{}}DL (U-Net variant)\end{tabular}
 &
  Radiologist &
  No &
  N/A & N/A
   \\ \hline
McGarry et al.~\citep{mcgarry2019gleason} &  \begin{tabular}[c]{@{}l@{}}TML (Radiomics,\\ Otsu thresholding)\end{tabular} &
   Pathologist &
  Yes &  RP &  \begin{tabular}[c]{@{}l@{}}Semi-automated \\ MRI-histopathology\\
  registration\end{tabular}\\ \hline
De Vente et al.~\citep{de2020deep} & DL  (U-Net variant) &
    \begin{tabular}[c]{@{}l@{}}Semi-automated region \\ growing from targeted \\ biopsy centroid\end{tabular} &
 Yes & \begin{tabular}[c]{@{}l@{}}Targeted\\  biopsy\end{tabular} &
  Biopsy-core coordinates
   \\ \hline
Sanyal et al.~\citep{sanyal2020automated} & DL (U-Net) &
  Radiologist & 
  Yes &
  \begin{tabular}[c]{@{}l@{}}Targeted\\  biopsy\end{tabular} &
  Pathology reports \\ \hline

Sumathipala et al.~\citep{sumathipala2018prostate} & DL (SPCNet variant)
 & Radiologist &
  Yes &
  \begin{tabular}[c]{@{}l@{}}RP and\\ targeted \\ biopsy\end{tabular} &
  \begin{tabular}[c]{@{}l@{}}Cognitive registration\\ or manually matching\end{tabular} \\ \hline
Cao et al.~\citep{cao2019joint} & DL (DeepLabV3+) &
  Radiologist &
  Yes &
  RP &
  \begin{tabular}[c]{@{}l@{}}Cognitive registration\\ or manually matching\end{tabular} \\ \hline
Bhattacharya et al.~\citep{bhattacharya2020corrsignet} & DL (SPCNet variant) &
  Pathologist & 
  Yes &
  RP &
  \begin{tabular}[c]{@{}l@{}}Automated\\ MRI-histopathology\\ registration\end{tabular} \\ \hline
\begin{tabular}[c]{@{}l@{}}Seetharaman et al.~\citep{seetharaman2021automated}\end{tabular} & DL (SPCNet) &
  Digital pathologist &
   Yes &
  RP &
  \begin{tabular}[c]{@{}l@{}}Automated\\ MRI-histopathology\\ registration\end{tabular} \\ \hline
\begin{tabular}[c]{@{}l@{}}Bhattacharya et al.~\citep{bhattacharya2021selective}\end{tabular} & DL (SPCNet variant) &
  Digital pathologist &
  Yes &
  RP &
  \begin{tabular}[c]{@{}l@{}}Automated\\ MRI-histopathology\\ registration\end{tabular} \\ \hline
\end{tabular}%
}
\end{table}

\section{Materials and Methods}
\subsection{Data Description}
All data for this IRB-approved retrospective chart review study was collected at Stanford University Medical Center. Two independent cohorts of subjects were used for this study. Cohort C1 was comprised of 115 patients who underwent radical prostatectomy, while cohort C2 included 275 men with or without prostate cancer who underwent MRI-guided targeted biopsie for PI-RADS scores $\geq$ 3 lesions. Subjects in cohort C1 had a pre-operative MRI prior to radical prostatectomy, and post-operative whole-mount histopathology images of the entire prostate. 
Subjects in cohort C2 had an MRI prior to biopsy which was used to guide the MRI-TRUS fusion biopsy procedure.

\subsubsection{MRI}
For subjects in both cohorts, multi-parametric MRI scans were acquired using 3.0T GE MRI scanners with surface coils and without an endorectal coil. Axial T2-weighted (T2w) MRI scans and Apparent Diffusion Coefficient (ADC) maps derived from Diffusion Weighted Images were used in this study (MRI acquisition characteristics detailed in Table 1 of Supplementary material).

\subsubsection{Histopathology Images}
For patients in cohort C1, the prostates removed via radical prostatectomy were sectioned into slices with the same thickness and in the same plane as the T2w scans, stained with Hematoxylin \& Eosin, and scanned into a digital format \citep{rusu2020registration, bhattacharya2021selective}. For patients in cohort C2, biopsy samples were stained with H\&E and subjected to pathological evaluation.

\textbf{Train-Test splits:} The machine learning models were trained using 75 patients from cohort C1 in a five-fold cross validation setting. The remaining 40 patients from cohort C1 and the entire cohort C2 (275 men) were used for independent testing of the models.

\subsection{Labels}

\subsubsection{Cancer and histologic grade labels}

\begin{figure*}[!ht]
    \centering
    \includegraphics[page=1, trim = 160 5 280 0, clip, width = 0.8\linewidth]{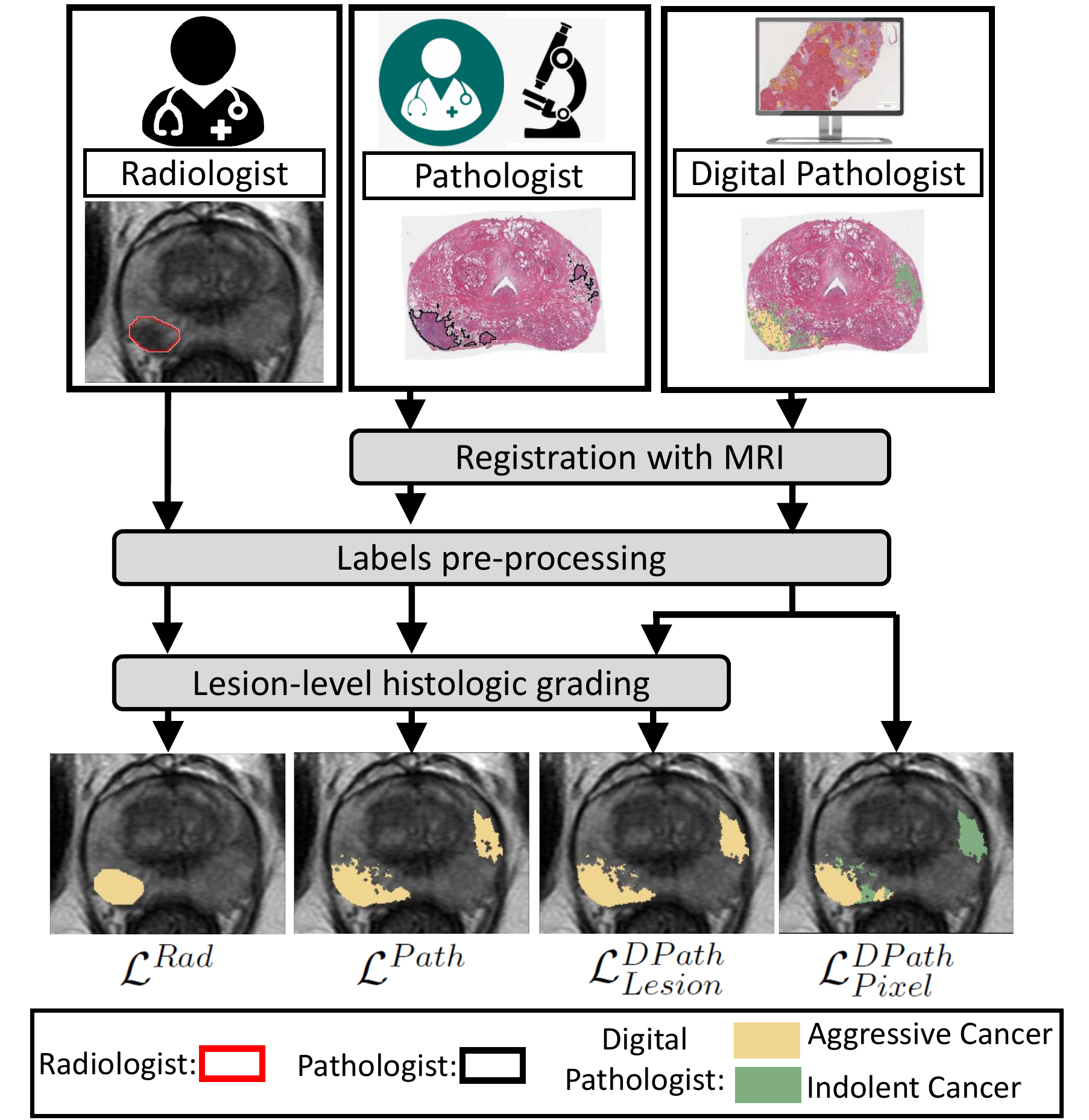} 
    \caption{Radiologists, pathologists or digital pathologists are used to create labels on MRI and serve to train deep learning models to detect cancer and aggressive cancer on MRI. The pathology labels ($\mathcal{L}^{Path}$, $\mathcal{L}^{DPath}_{Lesion}$ and $\mathcal{L}^{DPath}_{Pixel}$) are derived through annotations on whole-mount histopathology images and are mapped onto MRI through MRI-histopathology registration. The pixel-level digital pathologist label ($\mathcal{L}^{DPath}_{Pixel}$) enables identifying aggressive and indolent cancer components in mixed lesions, unlike the other label types.}
    \label{fig:flowchart_label}
\end{figure*}

\textbf{Cohort C1:} Patients in cohort C1 had four different types of cancer labels. Each label type annotated each pixel of the prostate into one of the three classes: (1) normal tissue, (2) indolent cancer, and (3) aggressive cancer.

A previously validated deep learning model on histopathology images (henceforth called the ``digital pathologist'') \citep{ryu2019automated} was used to predict Gleason patterns for each pixel of the prostate. Gleason pattern 3 predicted by the digital pathologist was considered indolent cancer, while Gleason patterns 4 and above were considered aggressive cancer. Regions of overlapping Gleason patterns 3 and 4 were considered aggressive cancer. 

\figurename~\ref{fig:flowchart_label} shows the flowchart for obtaining the different label types, described below:
\begin{enumerate}
    \item $\mathcal{L}^{Rad}$: Experienced radiologists outlined suspicious lesions on MR images prior to biopsy, and assigned PI-RADS scores to each lesion as part of routine clinical care. These radiologist-annotated lesions with PI-RADS scores $\geq$3 , after pathology confirmation were considered as $\mathcal{L}^{Rad}$ labels (\figurename~\ref{fig:label_comp_c1}c).
    
   Whole-mount histopathology specimens and histologic grade labels predicted by the digital pathologist \citep{ryu2019automated} on these specimens were used to confirm whether lesions outlined by radiologists corresponded to aggressive cancer (see ``pathology confirmation of radiologist labels'' below). The pixel-level Gleason patterns or histologic grade labels on histopathology images\citep{ryu2019automated} predicted by the digital pathologist were mapped onto pre-operative MRI using an MRI-histopathology registration \citep{rusu2020registration} platform (see Section \ref{sec:preprocess}). The digital pathologist predictions inside each radiologist annotation was used to derive pathology confirmations for that lesion. If a radiologist outline contained at least 1\% digital pathologist-predicted aggressive pixels, the annotation was considered as an aggressive lesion. If the radiologist outline had less than 1\% aggressive pixels, but had at least 1\% digital pathologist-predicted indolent pixels, it was considered as an indolent lesion. If a radiologist outline had less than 1\% aggressive or indolent pixels, it was considered as benign tissue. 
    
    \item $\mathcal{L}^{Path}$: An expert pathologist (C.A.K. with $>$ 10 years of experience) outlined the extent of cancer on whole-mount histopathology images. These pathologist annotations were converted to 3D lesions using morphological processing (see Section~\ref{sec:preprocess}). The digital pathologist-derived Gleason patterns \citep{ryu2019automated} were used to label each pathologist-annotated lesion into aggressive or indolent, in a way similar to the radiologist labels (at least 1\% aggressive pixels within the pathologist outline to be considered as an aggressive lesion). The pathologist labels were mapped onto pre-operative MRI using the MRI-histopathology registration platform \citep{rusu2020registration} (\figurename~\ref{fig:label_comp_c1}d).
       
    \item $\mathcal{L}^{DPath}_{Lesion}$: The pixel-level histologic grade labels from the digital pathologist were converted into lesion-level annotations through morphological processing (see Section~\ref{sec:preprocess}) and by considering the percentage of aggressive cancer pixels within a lesion outline, in a way similar to $\mathcal{L}^{Rad}$ and $\mathcal{L}^{Path}$. These lesion-level digital pathologist labels were then mapped onto MRI using the MRI-histopathology registration platform \citep{rusu2020registration} (\figurename~\ref{fig:label_comp_c1}e).
    
    \item $\mathcal{L}^{DPath}_{Pixel}$: The pixel-level histologic grade labels from the digital pathologist was used to derive pixel-level aggressive and indolent labels for the entire prostate (\figurename~\ref{fig:label_comp_c1}f). Unlike any other label type, pixel-level digital pathologist labels $\mathcal{L}^{DPath}_{Pixel}$ selectively labeled aggressive and indolent components of mixed lesions, instead of labeling the entire lesion as aggressive or indolent.  
     \end{enumerate}
\begin{table*}[!ht]
\vspace{0.5cm}
  
\centering
\begin{tabular}{|p{1.0in}||p{0.4in}|p{0.4in}|p{0.4in}|p{0.4in}||p{0.4in}|p{0.4in}|p{0.4in}|p{0.4in}||p{0.4in}|}
\hline 
Cohort &  \multicolumn{4}{c||}{C1-Train} & \multicolumn{4}{c||}{C1-Test} & C2 \\ \hline
Labels & $\mathcal{L}^{Rad}$ & $\mathcal{L}^{Path}$ & $\mathcal{L}^{DPath}_{Lesion}$ & $\mathcal{L}^{DPath}_{Pixel}$ & $\mathcal{L}^{Rad}$ & $\mathcal{L}^{Path}$ & $\mathcal{L}^{DPath}_{Lesion}$ & $\mathcal{L}^{DPath}_{Pixel}$ & $\mathcal{L}^{Rad}$ \\ \hline
\texttt{\#} of patients & 75 & 75 & 75 & 75 & 40 & 40 & 40 & 40 & 275\\ \hline
\texttt{\#} of patients with cancer & 75 & 75 & 75 & 75 & 40 & 40 & 40 & 40 & 160\\ \hline
\texttt{\#} of patients with labels & 71 & 75 & 75 & 75 & 31 & 40 & 40 & 40 & 160\\ \hline
\texttt{\#} of lesions  & 76 & 87 & 87 & 82 & 30 & 48 & 45 & 43 & 193\\ \hline
\texttt{\#} of aggressive lesions  & 63 & 82 & 83 & 49 & 25 & 44 & 44 & 31 & 132\\ \hline

\texttt{\#} of indolent lesions  & 13 & 5 & 4 & 33 & 5 & 4 & 1 & 12 & 61 \\ \hline
Lesion Volume($mm^3$) Mean (std)  & 2041 (3337) & 2559 (4575) & 2337 (3869) & 2408 (3907) & 1667 (1398) & 2223 (2736) & 2546 (2653) & 2612 (2633) & 1632 (2079) \\ \hline

\end{tabular}
\caption{Descriptive statistics of annotations from the different label types. Statistics for number of patients with labels are irrespective of lesion volume, whereas statistics for number of lesions are for lesions with volume $\geq$ 250mm$^3$.}
\label{tab:dataset-labels}
\end{table*}
     
\textbf{Pathology confirmation of radiologist labels:}\label{sec:path_conf}
Our study relied on the digital pathologist \citep{ryu2019automated} aggressive and indolent labels on whole mount histopathology images to provide pathology confirmation and type for the radiologist lesions in Cohort C1. Other prior studies \citep{de2020deep} have used histopathology information from targeted biopsy, yet we preferred the more accurate approach of using whole-mount images for pathology confirmation. Moreover, some of our patients lacked targeted biopsy information (i.e., systematic biopsy without lesion targeting or biopsies at outside institutions), further motivating the use of whole-mount histopathology images for pathology confirmation.

In order to study the concordance between pathology confirmation from targeted biopsy and the digital pathologist on whole mount histopathology images, we analyzed 69 patients in C1-train that had both targeted biopsy and digital pathologist confirmations. There were a total of 89 radiologist-annotated lesions in these 69 patients, and after pathology-confirmation these correspond to 67 of the $\mathcal{L}^{Rad}$ labels in cohort C1-train (Table \ref{tab:dataset-labels}). We found that the digital pathologist labels agreed with the targeted biopsy confirmations in 77.5\% (69/89) of the lesions. The digital pathologist upgraded 11.2\% (10/89) of the lesions (benign on targeted biopsy upgraded to indolent/aggressive cancer by digital pathologist, or indolent cancer on targeted biopsy upgraded to aggressive cancer by digital pathologist), and downgraded 11.2\% (10/89) of the lesions (indolent or aggressive on targeted biopsy downgraded to benign by digital pathologist, or aggressive on targeted biopsy downgraded to indolent or benign by digital pathologist). These upgrades could be due to sampling errors on targeted biopsy. Seven of the ten downgraded lesions had small proportions of cancer ($<$ 5\% cancerous tissue) or aggressive cancer ($\leq$ 15\% of Gleason pattern 4 or above in the cancerous tissue) in the targeted biopsy specimens, and small lesions ($<$ 250~mm$^3$ lesion volumes) outlined by pathologist and digital pathologists on whole-mount histopathology images. The remaining three downgrades were due to MRI-histopathology registration errors or missing histopathology tissue from the whole-mount specimens. Nonetheless, the digital pathologist labels provide a standardized approach for pathology confirmation of radiologist annotations in the absence of targeted biopsy information. The use of digital pathologist labels for pathology confirmation of radiologist annotations is also consistent with its use to label pathologist lesions into aggressive or indolent in this study.

\textbf{Cohort C2: }Patients in cohort C2 only had pathology-confirmed radiologist labels $\mathcal{L}^{Rad}$. Since all patients in cohort C2 had targeted biopsy at our institution, pathology-confirmation for the radiologist annotations in cohort C2 were derived from pathology of targeted biopsies. Radiologist lesions with targeted biopsy Gleason grade group$\geq$2 were considered as aggressive lesions, whereas lesions with targeted biopsy Gleason grade group of 1 were considered indolent lesions. Radiologist-annotated lesions whose targeted biopsies were benign, were considered as normal tissue. Table \ref{tab:dataset-labels} details the number of aggressive, indolent, and cancerous lesions with their mean volumes annotated by each label type in both cohorts.

\subsubsection{Prostate segmentations}
Prostate gland segmentations were available on all T2w MRI slices for all patients in both cohorts.
In addition, prostate gland segmentations were also available on all histopathology images of cohort C1. Prostate segmentations on all T2w slices were initially performed by medical students and trainees (with 6+ months experience in this task) and were carefully reviewed by our experts (C.A.K - a pathologist with 14 years experience, G.S. – a urologic oncologist with 13 years of experience, P.G. – a body MR imaging radiologist
with 14 years of experience, M.R. – an image analytics expert with 10 years of experience working on prostate cancer).

\subsection{Data Preprocessing} \label{sec:preprocess}

The data preprocessing was similar to our prior studies \citep{seetharaman2021automated, bhattacharya2021selective}, including (1) registration of the pre-operative MRI and post-operative histopathology images using the RAPSODI registration platform \citep{rusu2020registration} for cohort C1, (2) manual affine registrations between T2w and ADC images for cohort C1, (3) cropping and resampling to have the same pixel-size (0.29mm $\times$ 0.29mm) and the same X-Y dimensions (224 $\times$ 224) for both cohorts, (4) MRI intensity standardization \citep{nyul2000new, reinhold2019evaluating} and normalization for both cohorts (data preprocessing details in Section II of the Supplementary Material). 

The label preprocessing steps included forming lesions continuous in the MRI volume from pixel-level annotations using morphological closing and connected component analysis. The morphological closing operation was performed using a 3D structuring element formed by stacking 3 disks of sizes 0.5mm, 1.5mm, and 0.5mm. This structuring element was chosen to ensure that the generated lesions from pixel-level annotations faithfully represented the original annotations. Lesions with a volume less than 250 mm$^3$ were discarded from this study as these smaller lesions ($\approx$ 6mm $\times$ 6mm $\times$ 6mm) are unlikely to be seen on MRI, and have been considered as clinically insignificant in prior studies \cite{matoso2019defining, turkbey2019prostate}.

\subsection{Model Architectures}
Four different deep learning model architectures (SPCNet \citep{seetharaman2021automated}, U-Net \citep{schelb2019classification, saha2021end, ronneberger2015u, sanyal2020automated}, branched U-Net, and DeepLabv3+ \citep{cao2019joint}) were trained using each of the four label types. These four deep learning models were selected based on their previous performance in detecting and localizing prostate cancer (details of these architectures in Section III of the Supplementary material). All model architectures were evaluated to assess whether the effects of different labeling strategies were independent of the model architecture used.  Three consecutive slices of T2w-MRI and ADC images were used as inputs to all models, except for DeepLabv3+ which takes in a single slice of T2w and ADC images as input. All models were trained using a class-balanced cross-entropy loss function to enable multi-class prediction of each prostate pixel into one of the three classes: normal tissue, indolent cancer and aggressive cancer. A softmax activation function was used in the last layer of each model, and each prostate pixel was assigned the class with the maximum predicted probability. All models were trained in a five-fold cross-validation setting. No post-processing was done on the predicted labels. 

\subsection{Experimental Design}
The experimental design was setup to study the following: 
\subsubsection{Comparison between labeling strategies}

The different labels ($\mathcal{L}^{Rad}$, $\mathcal{L}^{Path}$, $\mathcal{L}^{DPath}_{Lesion}$, $\mathcal{L}^{DPath}_{Pixel}$) in cohort C1-test were analyzed with respect to each other in detecting and localizing cancer and aggressive cancer.  This analysis was done to study the concordance between the labels themselves, without any machine learning model-training.

\subsubsection{Establishing the best digital radiologist architecture} \label{sec:choose_dr}

Four different deep learning model architectures (SPCNet, U-Net, branched U-Net, DeepLabv3+) were trained on C1-train, each with the four different label types ($\mathcal{L}^{Rad}$, $\mathcal{L}^{Path}$, $\mathcal{L}^{DPath}_{Lesion}$, $\mathcal{L}^{DPath}_{Pixel}$), resulting in 16 different digital radiologists. Each model was trained in exactly the same way, with the same pre-processed data, class-balanced cross-entropy loss, batch size of 22, Adam optimizer and 30 training epochs. A learning rate of $10^{-4}$ was used for SPCNet and branched U-Net, $10^{-5}$ was used for U-Net and $10^{-3}$ was used for DeepLabv3+ architectures. These learning rates were chosen based on optimum performance in the validation set over a range of learning rates ($1\times10^{-5}$, $3 \times 10^{-5}$, $1 \times 10^{-4}$, $3 \times 10^{-4}$, $1 \times 10^{-3}$, $3 \times 10^{-3}$, $1 \times 10^{-2}$, $3 \times 10^{-2}$). The 16 different digital radiologist models were evaluated for the tasks of detecting cancer and aggressive cancer in cohorts C1-test, and in detecting cancer, aggressive cancer and indolent cancer in cohort C2 . The best digital radiologist model architecture was then chosen from the four different architectures (SPCNet, U-Net, branched U-Net, DeepLabv3+) based on their comparative evaluation.

\subsubsection{Studying the effect of different labeling strategies on digital radiologist performance}

The effect of the different label types on the performance of the digital radiologist they train was then studied by analyzing the performance of the best digital radiologist model architecture chosen in Section \ref{sec:choose_dr}

\subsection{Evaluation Methods}

The trained digital radiologist models were evaluated in cohort C1-test with respect to all four label types ($\mathcal{L}^{Rad}$, $\mathcal{L}^{Path}$, $\mathcal{L}^{DPath}_{Lesion}$, $\mathcal{L}^{DPath}_{Pixel}$). Evaluation in cohort C1-test generated 4 $\times$ 4 matrices for each evaluation metric, showing how a digital radiologist trained with one label type performed when evaluated with all the other label types. The trained digital radiologist models were also evaluated in cohort C2, which only had pathology-confirmed radiologist labels ($\mathcal{L}^{Rad}$). Evaluation in cohort C2 enabled studying generalizability of digital radiologists trained with different label types in an independent test set with different distribution of prostate cancer than cohort C1.

The digital radiologists were evaluated for their ability to detect and localize cancer (combined aggressive and indolent subtypes), aggressive cancer, and indolent cancer on prostate MRI on a lesion-level. For the lesion-level evaluation, a sextant-based approach was used \citep{seetharaman2021automated, bhattacharya2021selective}. True positives and false negatives were assessed using the ground truth and predicted labels, whereas true negatives and false positives were assessed by splitting the prostate into sextants, by first dividing it into left and right halves, and then dividing each half into 3 roughly equal regions (base, mid and apex) along the Z-axis. This sextant-based lesion-level evaluation is based upon how prostate biopsies are done in clinical practice, with two systematic biopsy cores from each sextant and additional targeted biopsies directed at the lesions. All evaluation was performed on a per-patient basis, and mean and standard deviation numbers for the entire test sets were reported. Lesion-level ROC-AUC, sensitivity, specificity and Dice coefficients were used as evaluation metrics (details of evaluation metrics reported in Section IV of Supplementary Material).

\section{Results}

Our comparison of different MR image-labeling approaches consisted of three parts. First, we compared the different labeling schemes to evaluate the accuracy of the radiologist and digital pathologist labels relative to the pathologist labels, irrespective of machine learning. Second, we compared multiple deep learning architectures to identify the one that performed best on the task of detecting prostate cancer and aggressive prostate cancer on MRI. Third, we carried out a thorough analysis of the performance of the best deep learning architecture in the context of the different labeling strategies. 
\subsection{Comparison between labeling strategies}

\begin{figure}[!ht]
    \centering
    \includegraphics[page=1, trim = 40 140 40 110, clip, width = \textwidth]{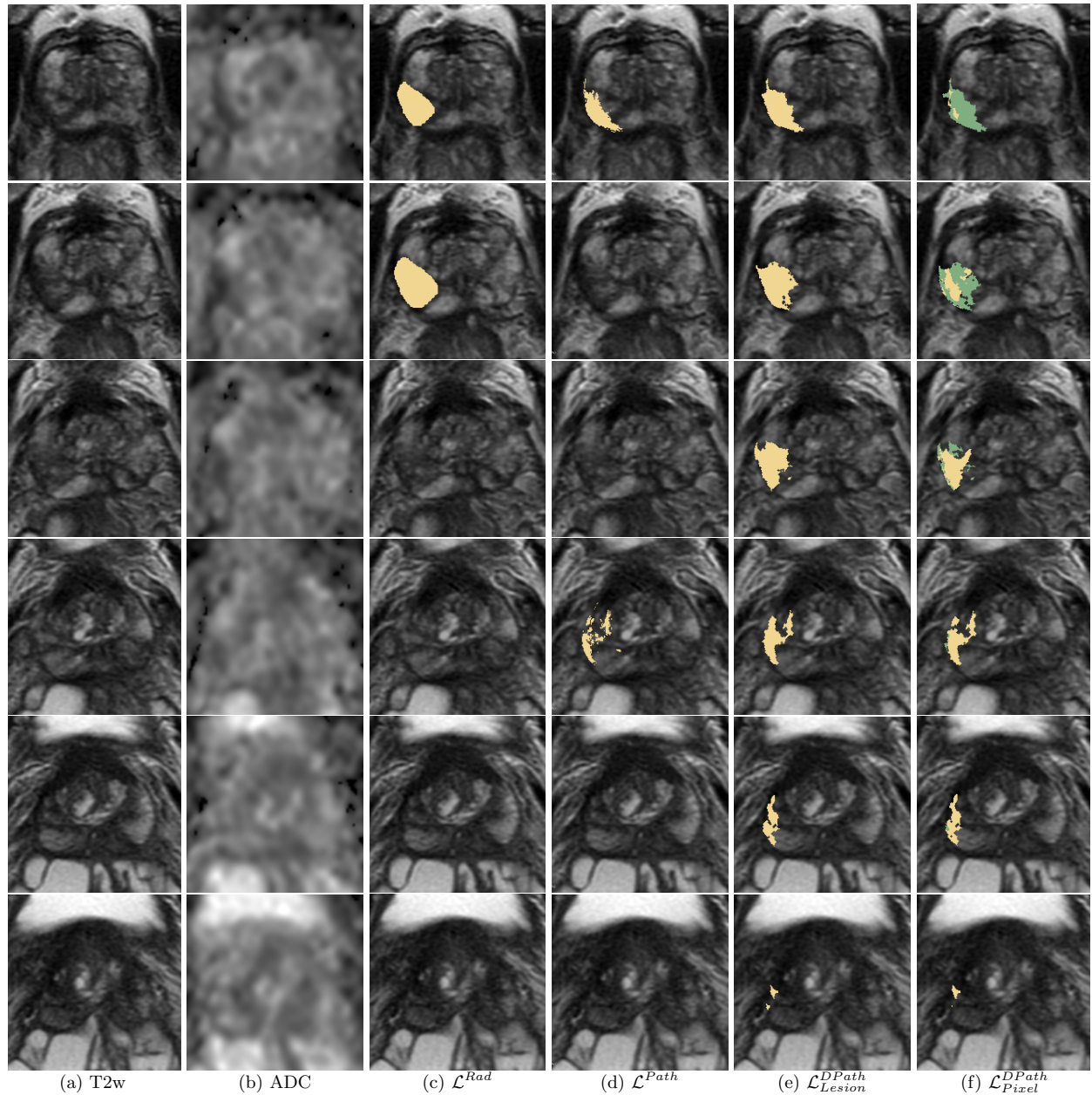}
    \caption{Differences in labeling strategies in a typical patient in cohort C1-test (aggressive cancer - yellow, indolent cancer - green)  showed on (a) T2w images and  (b) ADC images. The (c) radiologist labels ($\mathcal{L}^{Rad}$) and (d) pathologist labels ($\mathcal{L}^{Path}$) are present on some slices while the (e) lesion-level digital pathologist labels ($\mathcal{L}^{DPath}_{Lesion}$), and (f) pixel-level digital pathologist labels ($\mathcal{L}^{DPath}_{Pixel}$) exist on all slices. Digital pathologist labels strongly agree with pathologists while annotating aggressive and indolent cancer components in mixed lesions.
    }
    \label{fig:label_comp_c1}
\end{figure}

Annotating cancer extent on radiology or pathology images is tedious and rarely required for routine clinical care. Thus, for all practical purposes, for each patient, clinicians often outline cancerous lesions in some slices, e.g., slice with the larger extent, and skip the same lesion when it continues in other slices. Moreover, while radiologists and pathologists may outline the same lesions, they annotate the extent of the cancer differently. For example,
the radiologist annotated cancer on two slices (slices 1, 2 in \figurename~\ref{fig:label_comp_c1}c), while the pathologist outlined cancer on slices 1 and 4 (\figurename~\ref{fig:label_comp_c1}d) and skipped slices 2 and 3 due to time constraints and not because there are cancer-free. Unlike the radiologist and pathologist labels, the digital pathologist labels exist for all slices (\figurename~\ref{fig:label_comp_c1}e-f), while the pixel-level digital pathologist label ($\mathcal{L}^{DPath}_{Pixel}$) selectively identifies the aggressive (yellow) and indolent (green) cancer components in the mixed lesion. 
While differences exist between pathologist and digital pathologist labels, there is a strong agreement in cancer location and extent (\figurename~\ref{fig:label_comp_c1}). 


\begin{figure*}[!ht]
\centering
 \begin{subfigure}[b]{.48\linewidth}
  \centering
   \caption*{Cancer vs.~all}
  \end{subfigure}
   \begin{subfigure}[b]{.48\linewidth}
  \centering
   \caption*{Aggressive Cancer vs.~all}
  \end{subfigure}
\includegraphics[trim=0 0 5 40,clip,width = 0.24\textwidth]{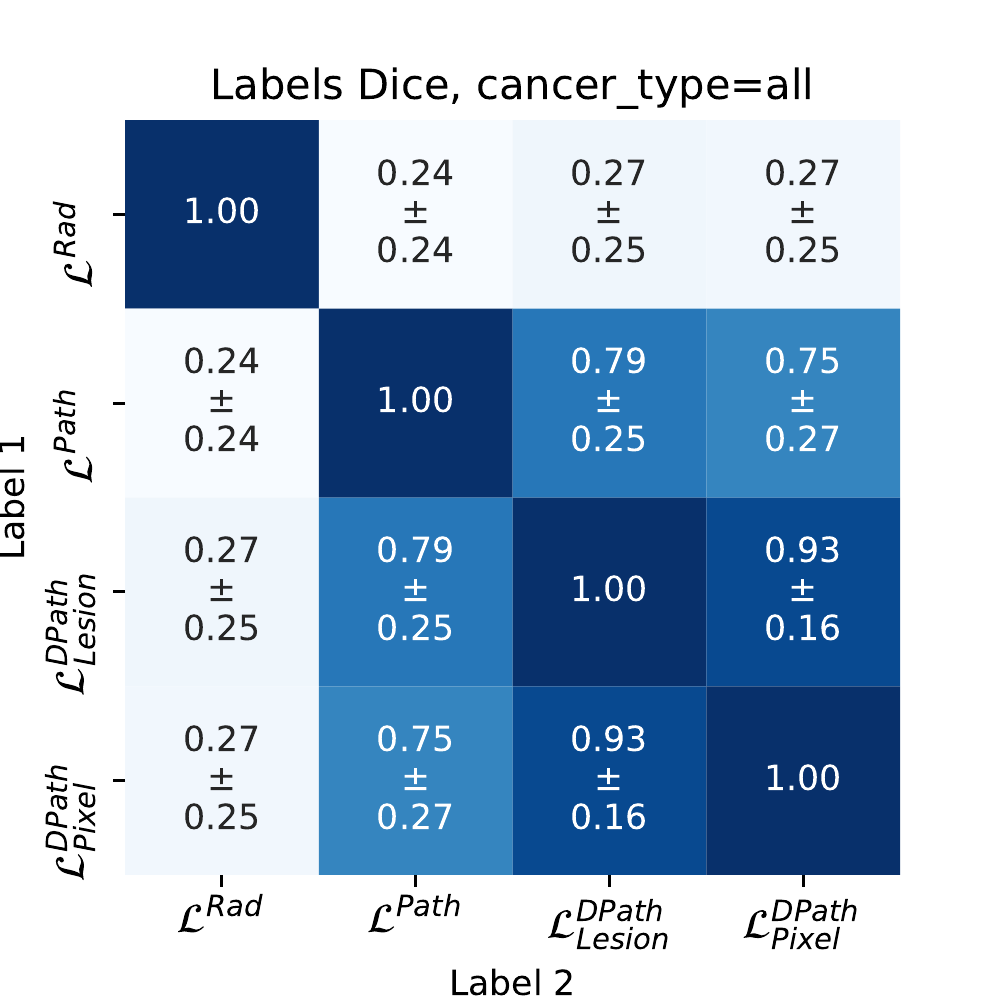}
\includegraphics[trim=0 0 5 40,clip,width = 0.24\textwidth]{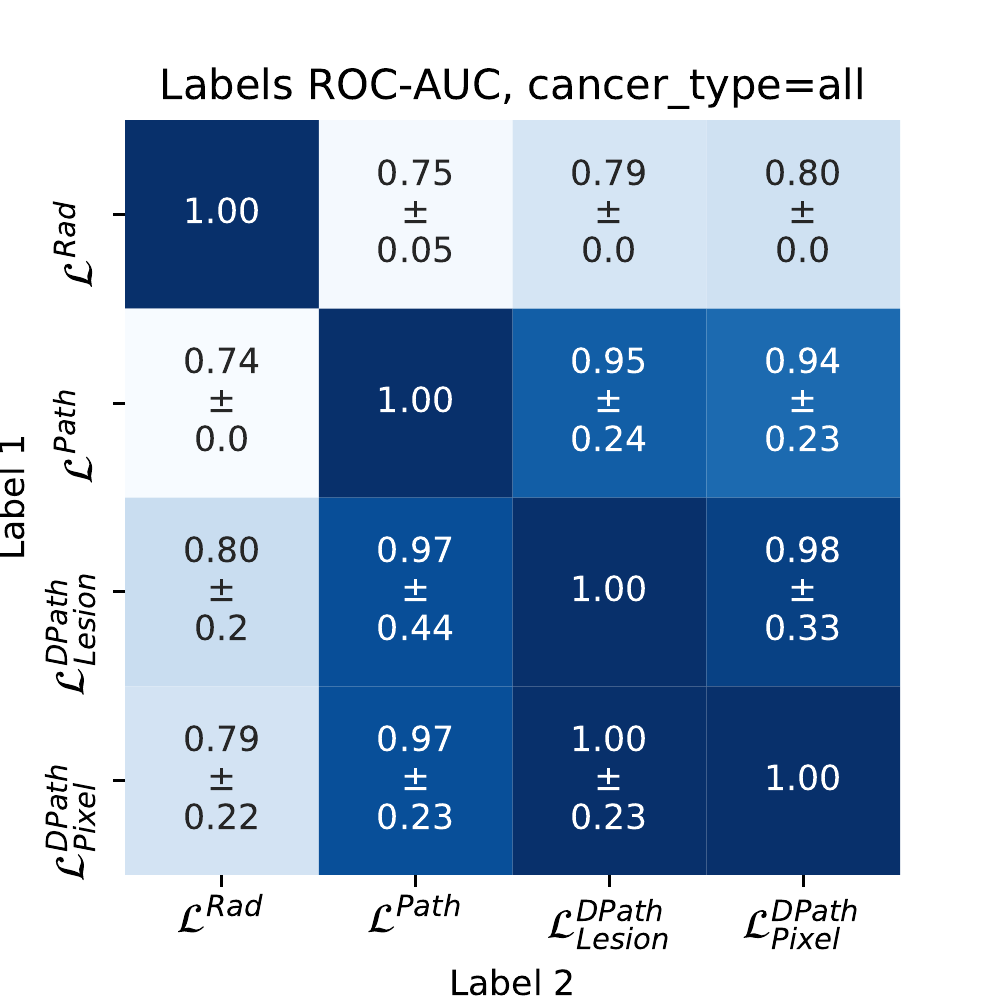}
\includegraphics[trim=0 0 5 40,clip,width = 0.24\textwidth]{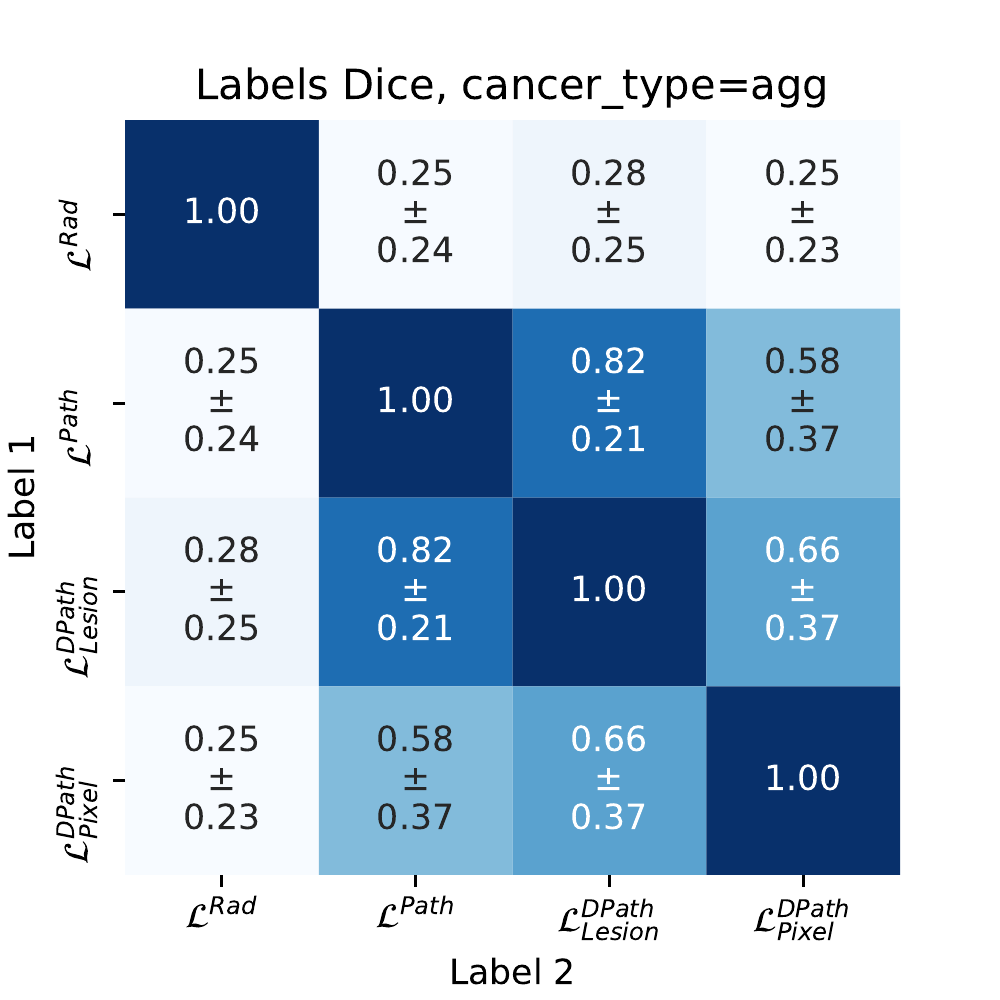}
\includegraphics[trim=0 0 5 40,clip,width = 0.24\textwidth]{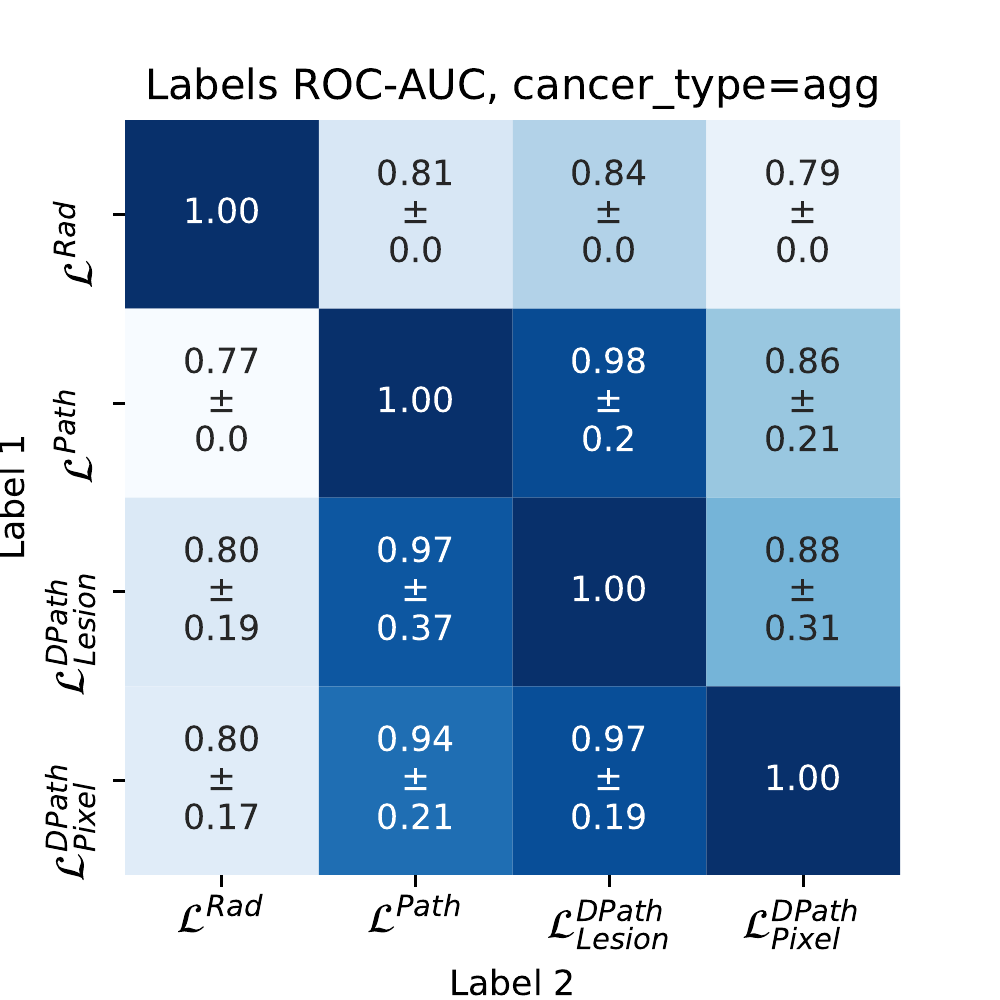}
\begin{subfigure}[b]{.24\linewidth}
  \centering
   \caption{Dice}
  \end{subfigure}
  \begin{subfigure}[b]{.24\linewidth}
  \centering
   \caption{Lesion ROC-AUC}
  \end{subfigure}
\begin{subfigure}[b]{.24\linewidth}
  \centering
   \caption{Dice}
  \end{subfigure}
\begin{subfigure}[b]{.24\linewidth}
  \centering
  \caption{Lesion ROC-AUC}
  \end{subfigure}
  
\caption{Quantitative comparison between cancer outlines of the different label types. (a) Dice overlap for cancer, (b) Lesion-level ROC-AUC for cancer, (c) Dice overlap for aggressive cancer, (d) Lesion-level ROC-AUC for aggressive cancer.}
     \label{fig:label_comp}
\end{figure*}

We quantitatively compared the label types for subjects in cohort C1-test using Dice similarity coefficient and  lesion level ROC-AUC (\figurename~\ref{fig:label_comp}). 
The radiologist labels ($\mathcal{L}^{Rad}$) measured a low Dice overlap (0.24 -- 0.28) and had a lesion-level ROC-AUCs ranging from 0.75 to 0.84 in cancer and aggressive cancer detection relative to pathology labels ($\mathcal{L}^{Path}$, $\mathcal{L}^{DPath}_{Lesion}$, $\mathcal{L}^{DPath}_{Pixel}$). These lower metrics of radiologist labels can be attributed to radiologists (1) not annotating cancer on all MRI slices, (2) underestimating cancer extents, and (3) missing MRI-invisible or hardly-visible lesions. Radiologist labels have lower lesion volumes than any kind of pathology labels, corresponding  to \textasciitilde 75\% of $\mathcal{L}^{Path}$ lesion-volumes, and \textasciitilde 65\% of $\mathcal{L}^{DPath}_{Lesion}$ lesion-volumes (Table \ref{tab:dataset-labels}). Moreover, 11\% of patients did not have any radiologist-outlined lesions but ended up having clinically significant cancer (Table \ref{tab:dataset-labels}). 
The radiologist labels were from the initial diagnostic read in the clinical care of the patients, essentially in vacuum, without any pathology information. Although this reflects the real-world scenario of routine clinical care, this also puts radiologists at an unfair disadvantage when comparing their initial diagnostic reads with post-operative surgical specimens. 

The lesion-level digital pathologist labels ($\mathcal{L}^{DPath}_{Lesion}$) achieved high (0.79-0.82) Dice overlap and very high agreement in lesion-level ROC-AUCs (cancer ROC-AUCs: 0.94-1.00; aggressive cancer ROC-AUCs: 0.86-0.97) with pathologist labels ($\mathcal{L}^{Path}$). While not perfect, the Dice overlaps can be attributed to the difference in resolution between the two kinds of pathologist labels, i.e., digital pathologists labeling each gland in detail, while it is impractical to annotate each gland on the whole-mount prostate histopathology images in detail by a human pathologist. Moreover, the pathologist may have not provided labels on all slices. 
    
The pixel-level digital pathologist labels ($\mathcal{L}^{DPath}_{Pixel}$) achieved high Dice overlaps with $\mathcal{L}^{Path}$ and $\mathcal{L}^{DPath}_{Lesion}$ for cancer, and achieved lower Dice overlaps (0.58$\pm$0.37, 0.66$\pm$0.37,)  with $\mathcal{L}^{Path}$ and $\mathcal{L}^{DPath}_{Lesion}$ for aggressive cancer. This low aggressive cancer Dice coefficient for $\mathcal{L}^{DPath}_{Pixel}$ is due to its selective labeling of aggressive and indolent cancer components in mixed cancerous lesions, unlike the other label types which label the entire lesion as aggressive or indolent.

\subsection{Establishing the best digital radiologist architecture}
\begin{table}[!h]
\centering
\caption{The SPCNet architecture achieved the best performance in detecting cancer and aggressive cancer in both cohorts irrespective of the label type used for training.}
\label{tab:model_comp_c2}
\resizebox{\textwidth}{!}{%
\begin{tabular}{|l|llll|llll|}
\hline
\hline
\multicolumn{9}{|c|}{\textbf{Cancer vs.~all}} \\
\hline
\hline
\multicolumn{9}{|c|}{Cohort C1-test (N = 40, number of lesions = 48). Evaluated against $\mathcal{L}^{Path}$.} \\
\hline
 &
  \multicolumn{4}{c|}{AUC-ROC} &
  \multicolumn{4}{c|}{Dice} \\ \hline
\multicolumn{1}{|c|}{\begin{tabular}[c]{@{}c@{}}Trained with \\ Label type\end{tabular}} &
  \multicolumn{1}{c}{SPCNet} &
  \multicolumn{1}{c}{U-Net} &
  \multicolumn{1}{c}{branched U-Net} &
  \multicolumn{1}{c|}{DeepLabv3+} &
  \multicolumn{1}{c}{SPCNet} &
  \multicolumn{1}{c}{U-Net} &
  \multicolumn{1}{c}{branched U-Net} &
  \multicolumn{1}{c|}{DeepLabv3+} \\ \hline
$\mathcal{L}^{Rad}$       & 0.87$\pm$0.22
   & \textbf{0.90$\pm$0.22}
   & 0.77$\pm$0.33
   & 0.88$\pm$0.21
   & \textbf{0.37$\pm$0.22}
   & 0.32$\pm$0.21
   & 0.31$\pm$0.22
   & 0.34$\pm$0.22
   \\ \hline
$\mathcal{L}^{Path}$      & \textbf{0.90$\pm$0.22}
   & 0.85$\pm$0.25
   & 0.82$\pm$0.32
   & 0.86$\pm$0.21
   & \textbf{0.39$\pm$0.19}
   & 0.33$\pm$0.17
   & 0.29$\pm$0.20
   & 0.32$\pm$0.23\\ \hline
$\mathcal{L}^{DPath}_{Lesion}$       &  \textbf{0.92$\pm$0.18}
   & 0.85$\pm$0.30
   & 0.89$\pm$0.24
   & 0.89$\pm$0.19
   &  \textbf{0.34$\pm$0.2}
   & 0.19$\pm$0.10
   & 0.28$\pm$0.20
   & 0.32$\pm$0.21
  \\ \hline
$\mathcal{L}^{DPath}_{Pixel}$    & \textbf{0.91$\pm$0.19}
   & 0.86$\pm$0.26
   & 0.83$\pm$0.27
   & \textbf{0.91$\pm$0.17}
   & \textbf{0.30$\pm$0.21}
   & \textbf{0.30$\pm$0.22}
   & 0.25$\pm$0.20
   &  \textbf{0.30$\pm$0.24}
   \\ \hline

\multicolumn{9}{|c|}{Cohort C2 (N = 160, number of lesions = 193).  Evaluated against $\mathcal{L}^{Rad}$.} \\
\hline
 &
  \multicolumn{4}{c|}{AUC-ROC} &
  \multicolumn{4}{c|}{Dice} \\ \hline
\multicolumn{1}{|c|}{\begin{tabular}[c]{@{}c@{}}Trained with \\ Label type\end{tabular}} &
  \multicolumn{1}{c}{SPCNet} &
  \multicolumn{1}{c}{U-Net} &
  \multicolumn{1}{c}{branched U-Net} &
  \multicolumn{1}{c|}{DeepLabv3+} &
  \multicolumn{1}{c}{SPCNet} &
  \multicolumn{1}{c}{U-Net} &
  \multicolumn{1}{c}{branched U-Net} &
  \multicolumn{1}{c|}{DeepLabv3+} \\ \hline
$\mathcal{L}^{Rad}$  
   &  \textbf{0.84$\pm$0.29}
   & 0.75$\pm$0.36
   & 0.82$\pm$0.33
   & 0.81$\pm$0.34
   & \textbf{0.39$\pm$0.28}
   & 0.35$\pm$0.24
   & 0.38$\pm$0.26
   & 0.39$\pm$0.27
\\ \hline
$\mathcal{L}^{Path}$      & \textbf{0.81$\pm$0.33}
   & 0.76$\pm$0.36
   & 0.78$\pm$0.34
   &  \textbf{0.81$\pm$0.32}
   &  \textbf{0.37$\pm$0.27}
   & 0.28$\pm$0.18
   & 0.36$\pm$0.25
   & 0.35$\pm$0.25
\\ \hline
$\mathcal{L}^{DPath}_{Lesion}$   &  \textbf{0.81$\pm$0.32}
   & 0.76$\pm$0.34
   & 0.77$\pm$0.35
   & 0.79$\pm$0.33
   &  \textbf{0.37$\pm$0.27}
   & 0.19$\pm$0.12
   & 0.35$\pm$0.26
   & 0.34$\pm$0.25
   \\ \hline
$\mathcal{L}^{DPath}_{Pixel}$      & \textbf{0.81$\pm$0.31}
   &0.81$\pm$0.31
   & 0.75$\pm$0.36
   & 0.80$\pm$0.33
   &  \textbf{0.35$\pm$0.29}
   & 0.34$\pm$0.22
   & 0.33$\pm$0.25
   &0.31$\pm$0.26
   \\ \hline

\hline
\multicolumn{9}{|c|}{\textbf{Aggressive Cancer vs.~all}} \\
\hline
\hline
\multicolumn{9}{|c|}{Cohort C1-test (N = 40, number of lesions = 44).  Evaluated against $\mathcal{L}^{Path}$.} \\
\hline
 &
  \multicolumn{4}{c|}{AUC-ROC} &
  \multicolumn{4}{c|}{Dice} \\ \hline
\multicolumn{1}{|c|}{\begin{tabular}[c]{@{}c@{}}Trained with \\ Label type\end{tabular}} &
  \multicolumn{1}{c}{SPCNet} &
  \multicolumn{1}{c}{U-Net} &
  \multicolumn{1}{c}{branched U-Net} &
  \multicolumn{1}{c|}{DeepLabv3+} &
  \multicolumn{1}{c}{SPCNet} &
  \multicolumn{1}{c}{U-Net} &
  \multicolumn{1}{c}{branched U-Net} &
  \multicolumn{1}{c|}{DeepLabv3+} \\ \hline
$\mathcal{L}^{Rad}$      & 0.88$\pm$0.24
   & \textbf{0.91$\pm$0.23}
   & 0.78$\pm$0.32
   & \textbf{0.91$\pm$0.20}
   & \textbf{0.36$\pm$0.39}
   & 0.31$\pm$0.21
   & 0.31$\pm$0.22
   & 0.34$\pm$0.22\\ \hline
$\mathcal{L}^{Path}$   &  \textbf{0.91$\pm$0.21}
   & 0.88$\pm$0.25
   & 0.83$\pm$0.30
   & 0.90$\pm$0.19
   & \textbf{0.39$\pm$0.19}
   & 0.32$\pm$0.17
   & 0.29$\pm$0.20
   & 0.33$\pm$0.23\\ \hline
$\mathcal{L}^{DPath}_{Lesion}$    &\textbf{0.92$\pm$0.19}
   & 0.85$\pm$0.31
   & 0.90$\pm$0.23
   & \textbf{0.92$\pm$0.17}
   & \textbf{0.34$\pm$0.20}
   & 0.18$\pm$0.10
   & 0.28$\pm$0.21
   & 0.33$\pm$0.21
  \\ \hline
$\mathcal{L}^{DPath}_{Pixel}$         & 0.91$\pm$0.19
   & 0.90$\pm$0.20
   & 0.86$\pm$0.26
   & \textbf{0.92$\pm$0.16}
   & \textbf{0.31$\pm$0.21}
   & 0.30$\pm$0.22
   & 0.25$\pm$0.20
   & \textbf{0.31$\pm$0.24}
   \\ \hline

\multicolumn{9}{|c|}{Cohort C2 (N = 160, number of lesions = 132).  Evaluated against $\mathcal{L}^{Rad}$.} \\
\hline
 &
  \multicolumn{4}{c|}{AUC-ROC} &
  \multicolumn{4}{c|}{Dice} \\ \hline
\multicolumn{1}{|c|}{\begin{tabular}[c]{@{}c@{}}Trained with \\ Label type\end{tabular}} &
  \multicolumn{1}{c}{SPCNet} &
  \multicolumn{1}{c}{U-Net} &
  \multicolumn{1}{c}{branched U-Net} &
  \multicolumn{1}{c|}{DeepLabv3+} &
  \multicolumn{1}{c}{SPCNet} &
  \multicolumn{1}{c}{U-Net} &
  \multicolumn{1}{c}{branched U-Net} &
  \multicolumn{1}{c|}{DeepLabv3+} \\ \hline
$\mathcal{L}^{Rad}$ & \textbf{0.89$\pm$0.24}
   & 0.80$\pm$0.33
   & 0.86$\pm$0.30
   & 0.86$\pm$0.30
   & 0.43$\pm$0.26
   & 0.38$\pm$0.23
   & 0.42$\pm$0.24
   & \textbf{0.44$\pm$0.24} \\ \hline
$\mathcal{L}^{Path}$    & \textbf{0.87$\pm$0.27}
   & 0.83$\pm$0.31
   & 0.85$\pm$0.30
   & 0.86$\pm$0.27
   & \textbf{0.41$\pm$0.25}
   & 0.30$\pm$0.18
   & 0.40$\pm$0.23
   & 0.39$\pm$0.24\\ \hline
$\mathcal{L}^{DPath}_{Lesion}$  &  \textbf{0.87$\pm$0.26}
   & 0.81$\pm$0.32
   & 0.83$\pm$0.23
   & 0.86$\pm$0.28
   &  \textbf{0.42$\pm$0.25}
   & 0.20$\pm$0.11
   & 0.39$\pm$0.24
   & 0.39$\pm$0.25 \\ \hline
$\mathcal{L}^{DPath}_{Pixel}$
   & \textbf{0.88$\pm$0.27}
   & 0.86$\pm$0.27
   & 0.80$\pm$0.33
   & 0.85$\pm$0.31
   & \textbf{0.40$\pm$0.28}
   & 0.38$\pm$0.21
   & 0.36$\pm$0.24
   & 0.37$\pm$0.26
   \\ \hline
\end{tabular}%
}
\end{table}
We compared the four architectures (SPCNet, U-Net, branched U-Net, DeepLabv3+) trained with different label types in detecting and localizing cancer and aggressive cancer on a lesion-level (Table \ref{tab:model_comp_c2}). In cohort C1-test, models trained were evaluated with respect to pathologist labels ($\mathcal{L}^{Path}$), while in cohort C2, they were evaluated with respect to biopsy-confirmed radiologist labels ($\mathcal{L}^{Rad}$). 
SPCNet outperformed other models in most metrics and most evaluation types and thereby was chosen as the optimum digital radiologist for analyzing the effect of the different label types in the subsequent sections.

\subsection{Studying the effect of different labeling strategies on digital radiologist performance}

\begin{figure}[!h]
    \centering
    \includegraphics[page=3, trim = 40 140 40 120, clip, width = \textwidth]{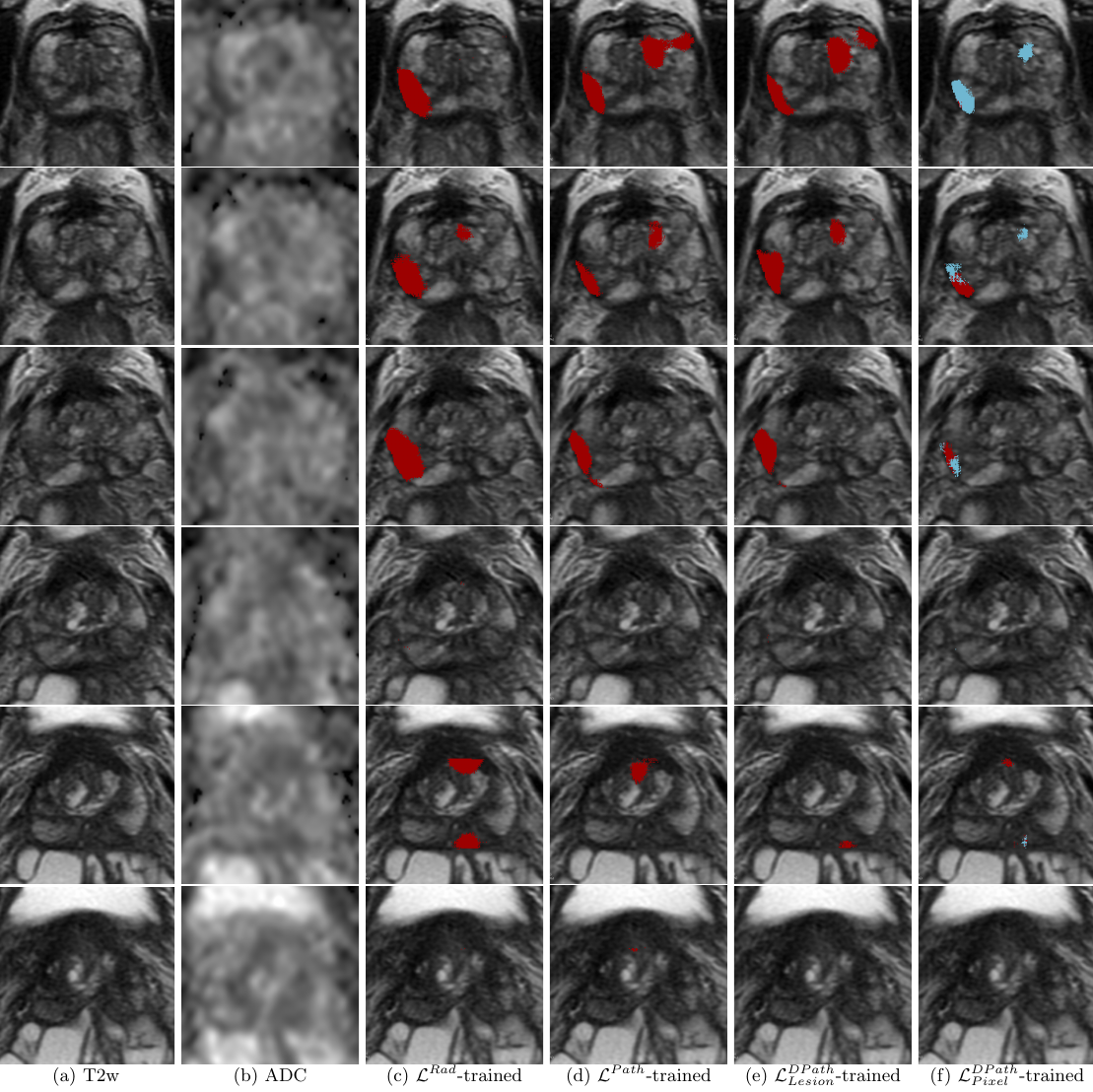}
    \caption{Predictions from SPCNet trained with different label types of a typical patient from cohort C1-test (same as \figurename~\ref{fig:label_comp_c1}) show that only $\mathcal{L}^{DPath}_{Pixel}$-trained SPCNet (f) selectively identified the aggressive and indolent cancer components in the lesion, while all other models detected the lesion as aggressive (SPCNet predictions: aggressive cancer (red), indolent cancer (blue)). (a) T2w images, (b) ADC images, (c) $\mathcal{L}^{Rad}$-trained SPCNet predictions, (d) $\mathcal{L}^{Path}$-trained SPCNet predictions, (e) $\mathcal{L}^{DPath}_{Lesion}$-trained SPCNet predictions, (f) $\mathcal{L}^{DPath}_{Pixel}$-trained SPCNet predictions.}
    \label{fig:spcnet_multislice_c1}
\end{figure}

\begin{figure}[!h]
    \centering
    \includegraphics[page=1, trim = 50 170 50 120, clip, width = \textwidth]{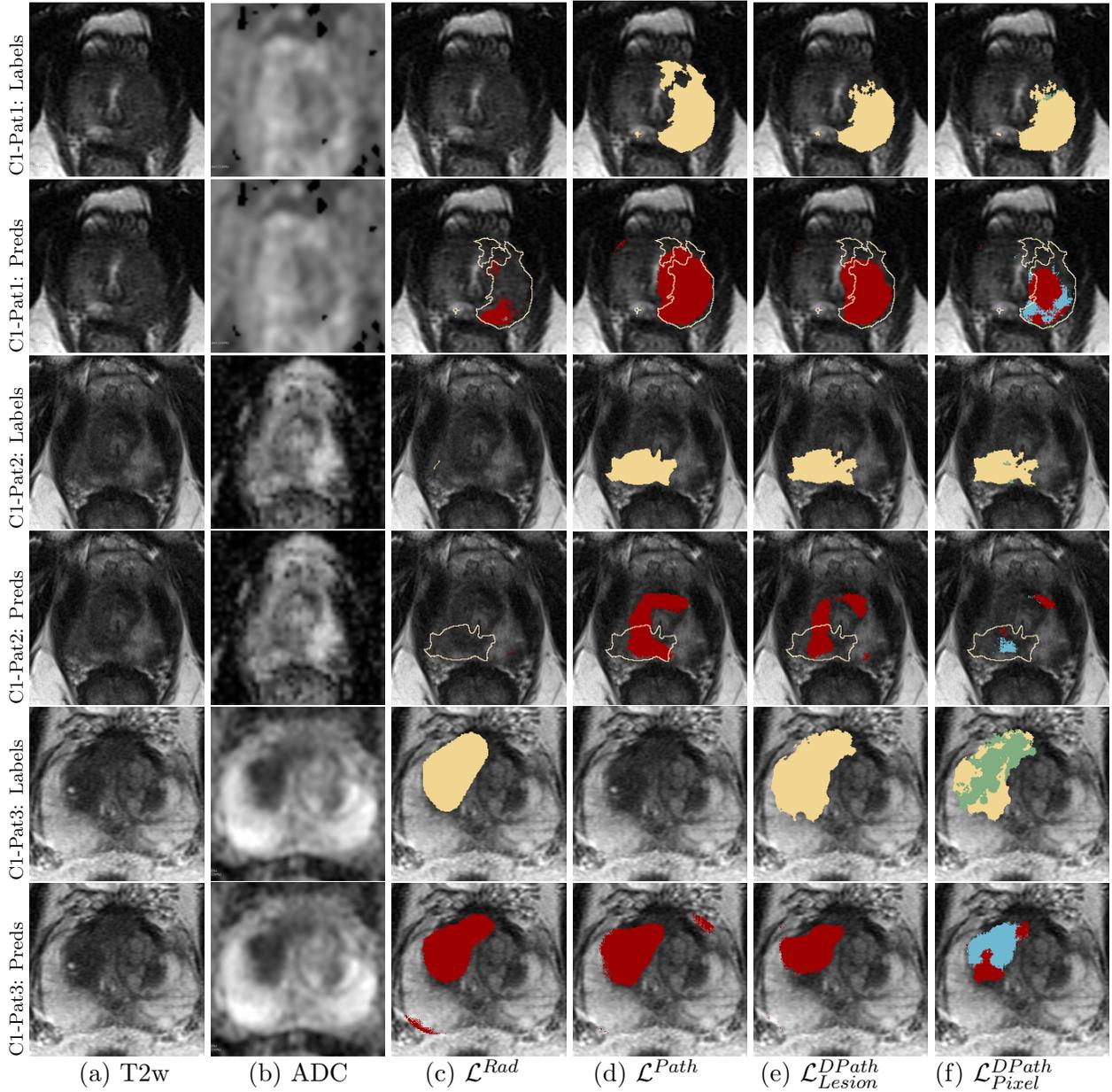}
    
     \begin{subfigure}[b]{.15\linewidth}
  \centering
   \caption{T2w}
  \end{subfigure}
\begin{subfigure}[b]{.15\linewidth}
  \centering
  \caption{ADC}
  \end{subfigure}
\begin{subfigure}[b]{.15\linewidth}
  \centering
  \caption{$\mathcal{L}^{Rad}$}
  \end{subfigure}  
\begin{subfigure}[b]{.15\linewidth}
  \centering
  \caption{$\mathcal{L}^{Path}$}
  \end{subfigure}  
  \begin{subfigure}[b]{.15\linewidth}
  \centering
  \caption{$\mathcal{L}^{DPath}_{Lesion}$}
  \end{subfigure}  
 \begin{subfigure}[b]{.15\linewidth}
  \centering
  \caption{$\mathcal{L}^{DPath}_{Pixel}$}
  \end{subfigure}

    \caption{Labels and SPCNet predictions for three different patients from cohort C1-test (Labels: aggressive cancer (yellow), indolent cancer (green)); SPCNet predictions: aggressive cancer (red), indolent cancer (blue)) on (a) T2w and (b) ADC images. The (c) $\mathcal{L}^{Rad}$ labels and $\mathcal{L}^{Rad}$-trained SPCNet predictions may miss cancers or underestimate cancer extent. The (d) $\mathcal{L}^{Path}$ labels and $\mathcal{L}^{Path}$-trained SPCNet predictions, and the (e)  $\mathcal{L}^{DPath}_{Lesion}$ and $\mathcal{L}^{DPath}_{Lesion}$-trained SPCNet predictions show strong agreement in cancer localization and extent. The (f) $\mathcal{L}^{DPath}_{Pixel}$ and $\mathcal{L}^{DPath}_{Pixel}$-trained SPCNet predictions can selectively identify and localize the aggressive and indolent cancer components in the mixed lesions unlike any other label or prediction type. The outline for columns with SPCNet predictions correspond to pathologist annotations. }
    \label{fig:spcnet_multipat_c1}
\end{figure}

 \begin{figure}[!h]
    \centering
    \includegraphics[page=2, trim = 50 323 50 290, clip, width = \textwidth]{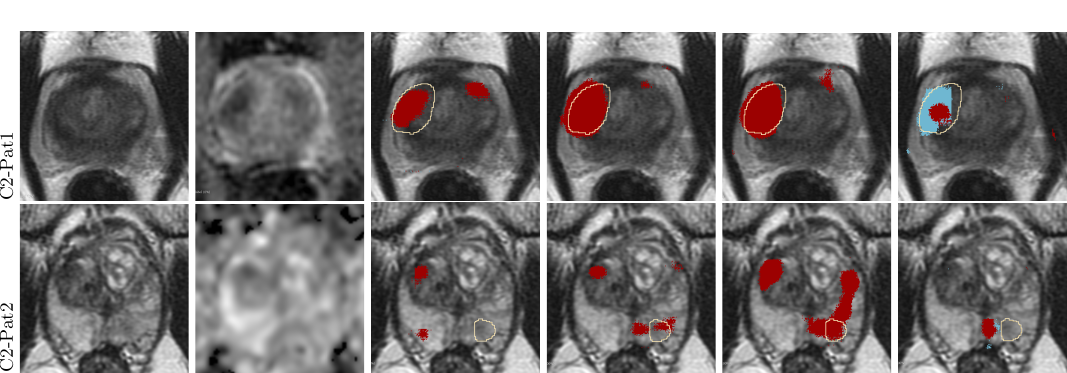}
 
 \begin{subfigure}[b]{.15\linewidth}
  \centering
   \caption{T2w}
  \end{subfigure}
\begin{subfigure}[b]{.15\linewidth}
  \centering
  \caption{ADC}
  \end{subfigure}
\begin{subfigure}[b]{.15\linewidth}
  \centering
  \caption{$\mathcal{L}^{Rad}$}
  \end{subfigure}  
\begin{subfigure}[b]{.15\linewidth}
  \centering
  \caption{$\mathcal{L}^{Path}$}
  \end{subfigure}  
  \begin{subfigure}[b]{.15\linewidth}
  \centering
  \caption{$\mathcal{L}^{DPath}_{Lesion}$}
  \end{subfigure}  
 \begin{subfigure}[b]{.15\linewidth}
  \centering
  \caption{$\mathcal{L}^{DPath}_{Pixel}$}
  \end{subfigure}

    \caption{SPCNet predictions for two different patients from cohort C2 on (a) T2w and (b) ADC images. The (c)$\mathcal{L}^{Rad}$-trained SPCNet predictions miss the cancer in the row 2 patient C2-Pat2. The (d)$\mathcal{L}^{Path}$-trained and (e) $\mathcal{L}^{DPath}_{Lesion}$-trained SPCNet predictions detect the lesions in both patients, with the (e) $\mathcal{L}^{DPath}_{Lesion}$-trained predictions having the highest overlap with the cancer extent. The (f) $\mathcal{L}^{DPath}_{Pixel}$-trained SPCNet predictions are slightly off from the $\mathcal{L}^{Rad}$ labels for the row 2 patient C2-Pat2. The outlines for columns with SPCNet-predictions correspond to radiologist labels ($\mathcal{L}^{Rad}$).}
    \label{fig:spcnet_multipat_c2}
\end{figure}

\subsubsection{Qualitative comparison}

Digital radiologists trained with radiologist labels ($\mathcal{L}^{Rad})$ could detect cancer in both cohorts (Figures~\ref{fig:spcnet_multislice_c1}c, \ref{fig:spcnet_multipat_c1}c and \ref{fig:spcnet_multipat_c2}c), but in comparison with other digital radiologists they missed some cancers (\figurename~\ref{fig:spcnet_multipat_c1}c, row 4, C1-Pat2:Preds, and \figurename ~\ref{fig:spcnet_multipat_c2}c, row 2, C2-Pat2), and underestimated cancer extent in some patients (\figurename~\ref{fig:spcnet_multipat_c1}c, row2, C1-Pat1:Preds and \figurename~\ref{fig:spcnet_multipat_c2}c, row 2, C2-Pat1).

Digital radiologists trained with lesion-level pathology labels ($\mathcal{L}^{Path}$ and $\mathcal{L}^{DPath}_{Lesion}$) had the best (and very similar) performances in detecting and localizing cancer, and also in capturing the true extent of the cancer (Figures~\ref{fig:spcnet_multislice_c1}, \ref{fig:spcnet_multipat_c1} and \ref{fig:spcnet_multipat_c2}, columns d and e). Digital radiologists trained with pixel-level digital pathologist labels ($\mathcal{L}^{DPath}_{Pixel}$) are the only ones to selectively identify aggressive and indolent cancer in mixed lesions (\figurename~\ref{fig:spcnet_multislice_c1}f and \figurename~\ref{fig:spcnet_multipat_c1}f, row 6, C1-Pat3: Preds), albeit  sometimes having less cancer extent than the $\mathcal{L}^{Path}$ and $\mathcal{L}^{DPath}_{Lesion}$-trained digital radiologists (\figurename~\ref{fig:spcnet_multipat_c1}f, row 4, C1-Pat2: Preds). Predictions from the $\mathcal{L}^{DPath}_{Pixel}$-trained digital radiologist for the row 2 patient (C2-Pat2) is slightly off from the actual ground truth lesion annotation.
 
\subsubsection{Quantitative comparison}
\begin{figure*}[!h]
\centering
\caption*{Cancer vs.~all}
\includegraphics[trim=0 0 5 31,clip,width = 0.24\textwidth]{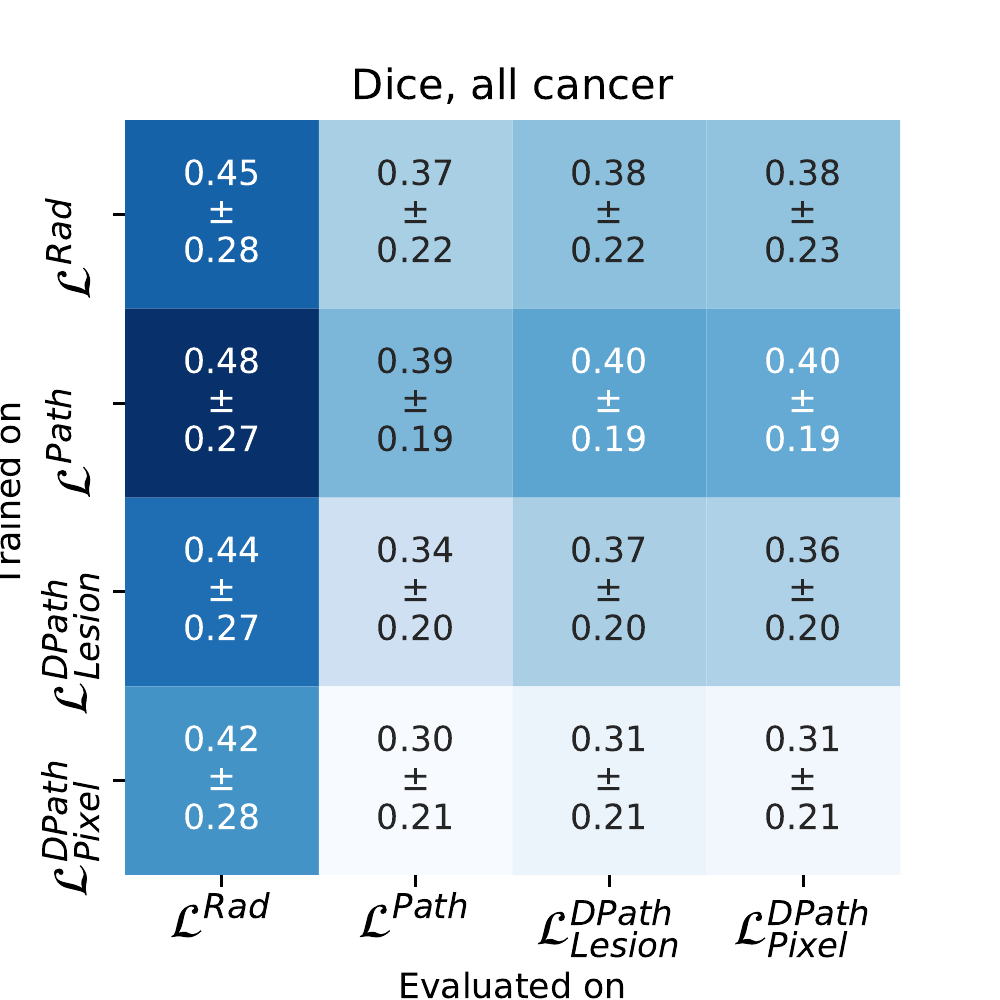}
\includegraphics[trim=0 0 5 31,clip,width = 0.24\textwidth]{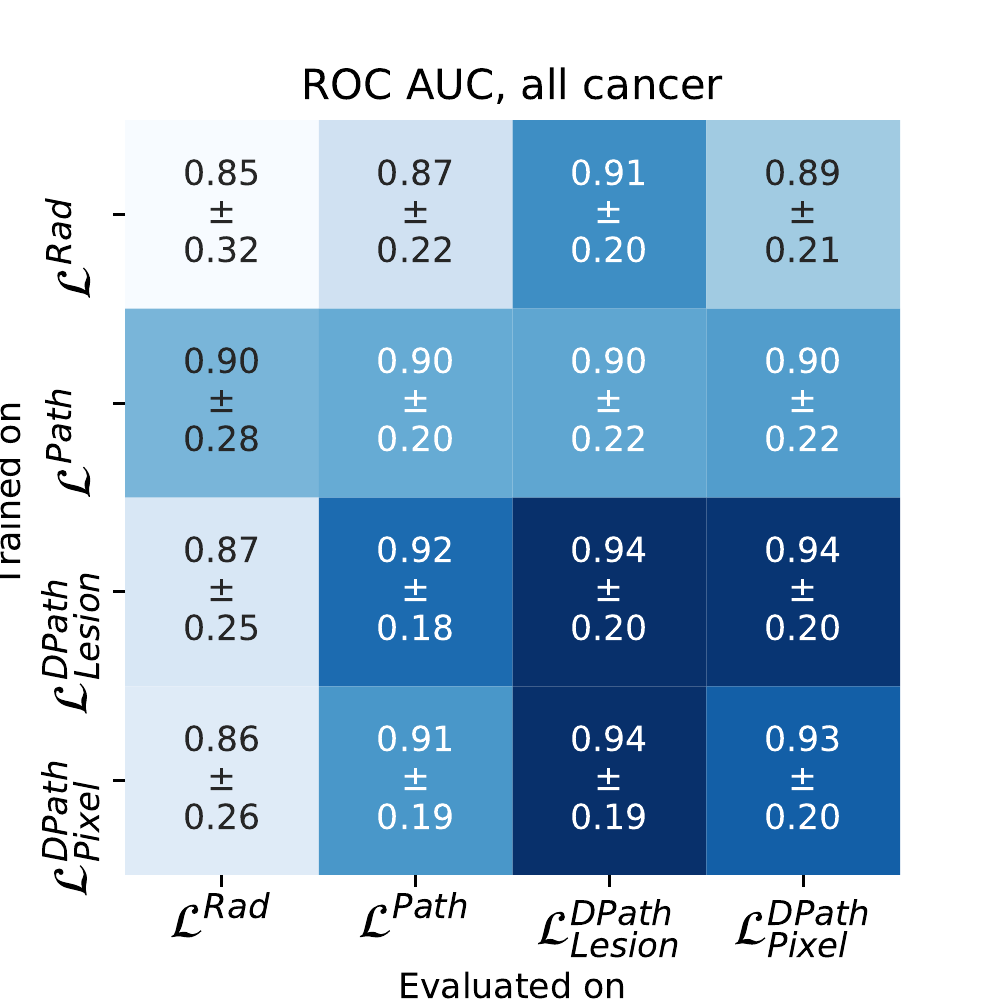}
\includegraphics[trim=0 0 5 31,clip,width = 0.24\textwidth]{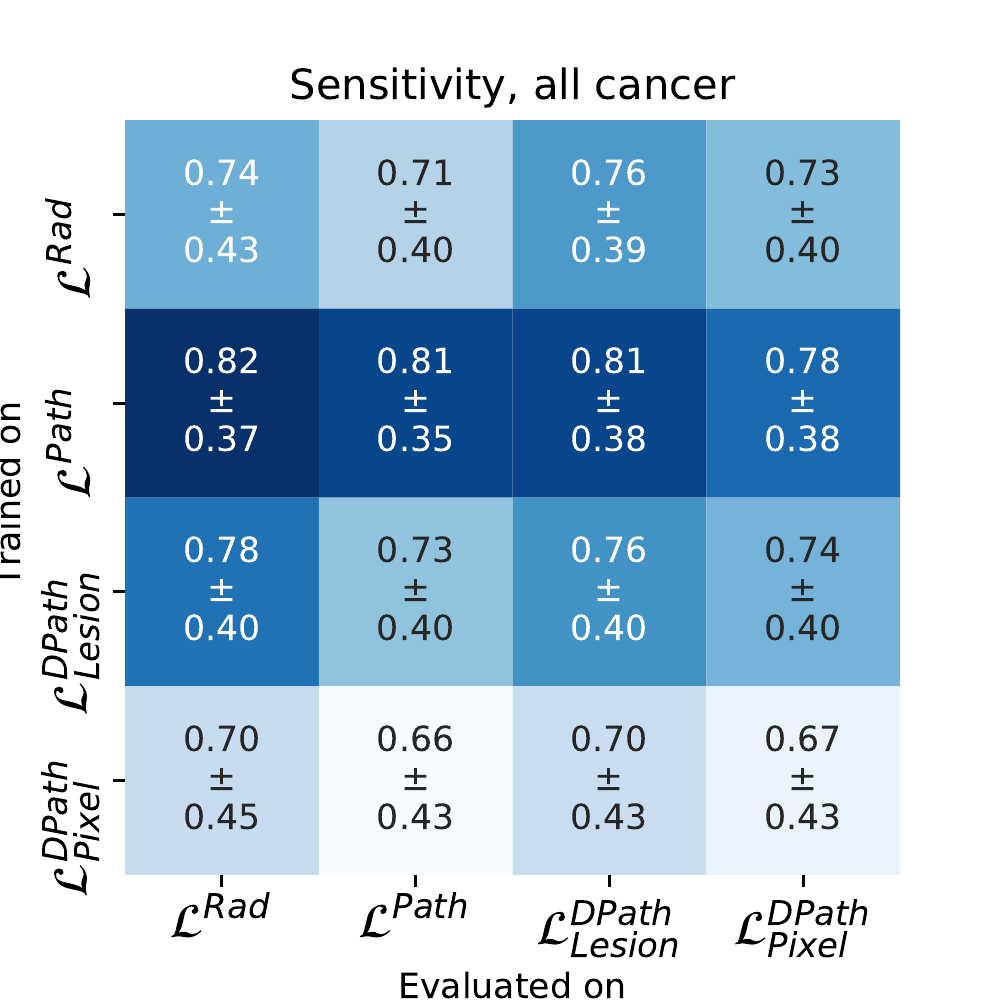}
\includegraphics[trim=0 0 5 31,clip,width = 0.24\textwidth]{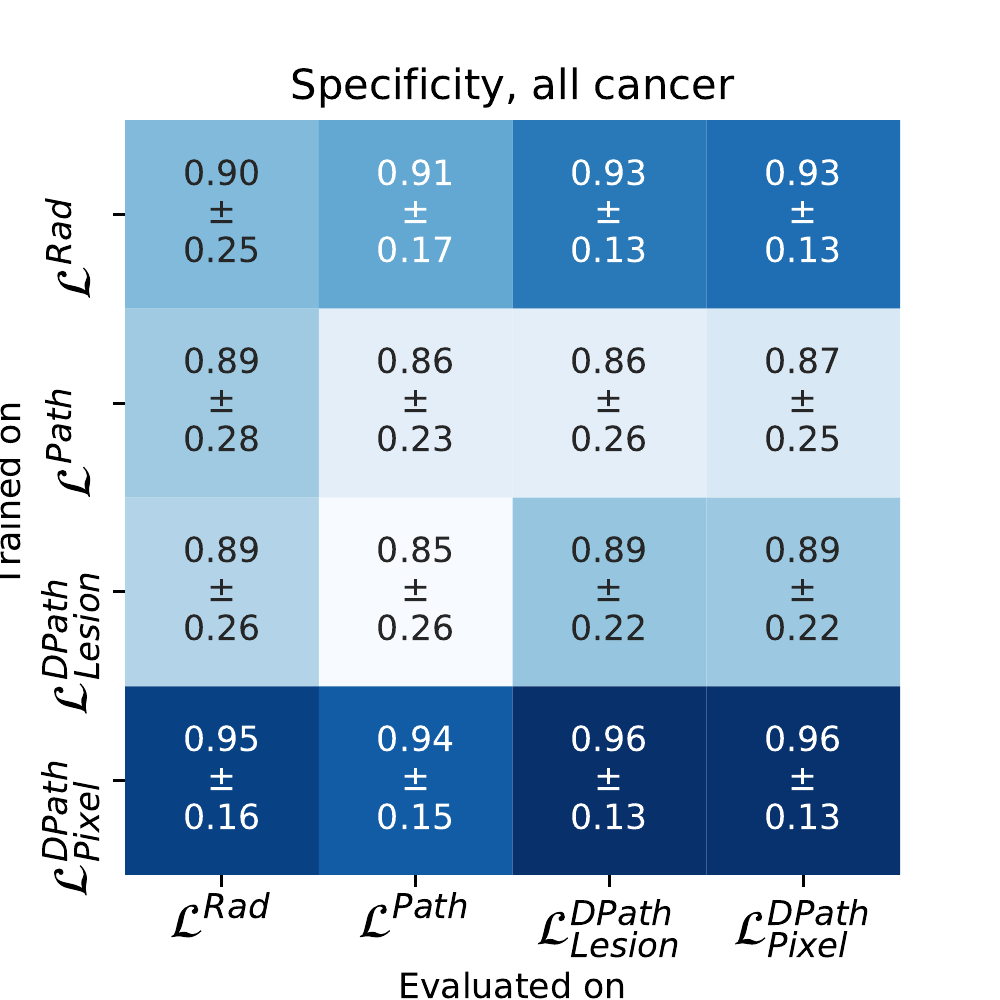}

\caption*{Aggressive Cancer vs.~all}
\includegraphics[trim=0 0 5 31,clip,width = 0.24\textwidth]{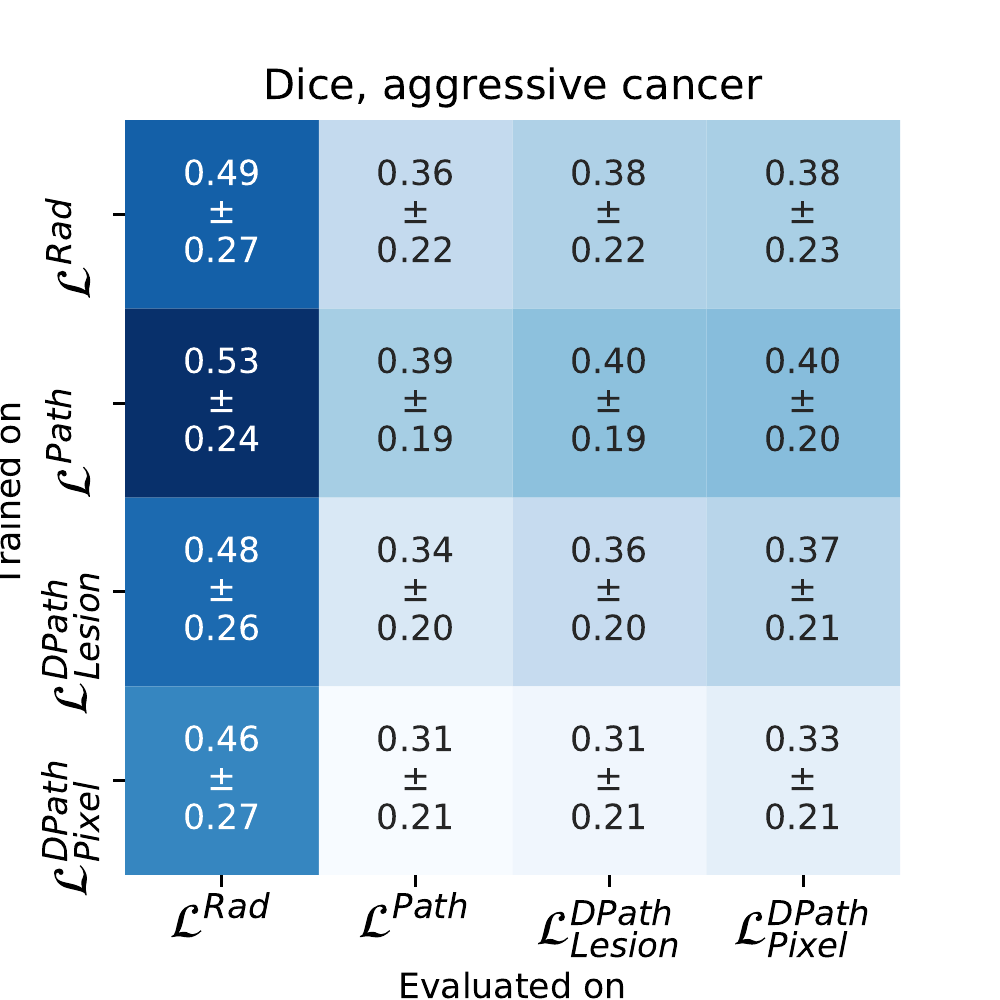}
\includegraphics[trim=0 0 5 31,clip,width = 0.24\textwidth]{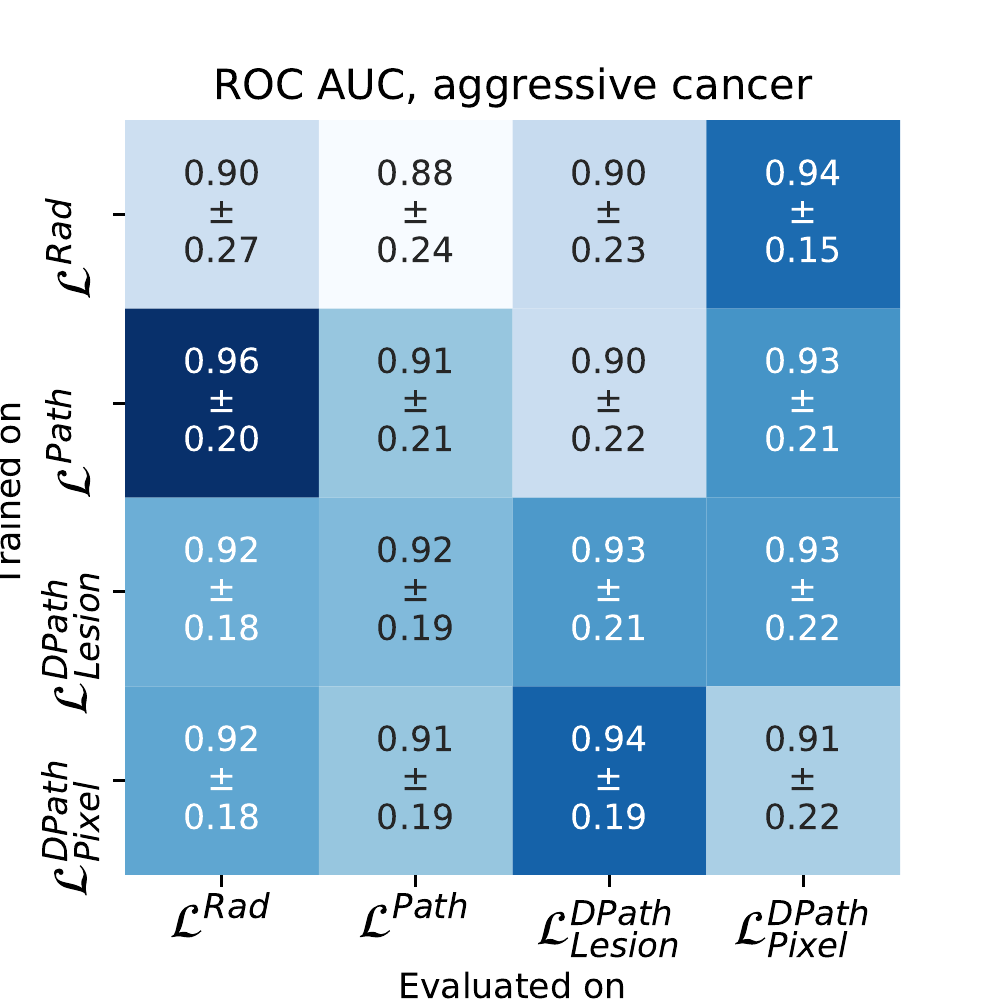}
\includegraphics[trim=0 0 5 31,clip,width = 0.24\textwidth]{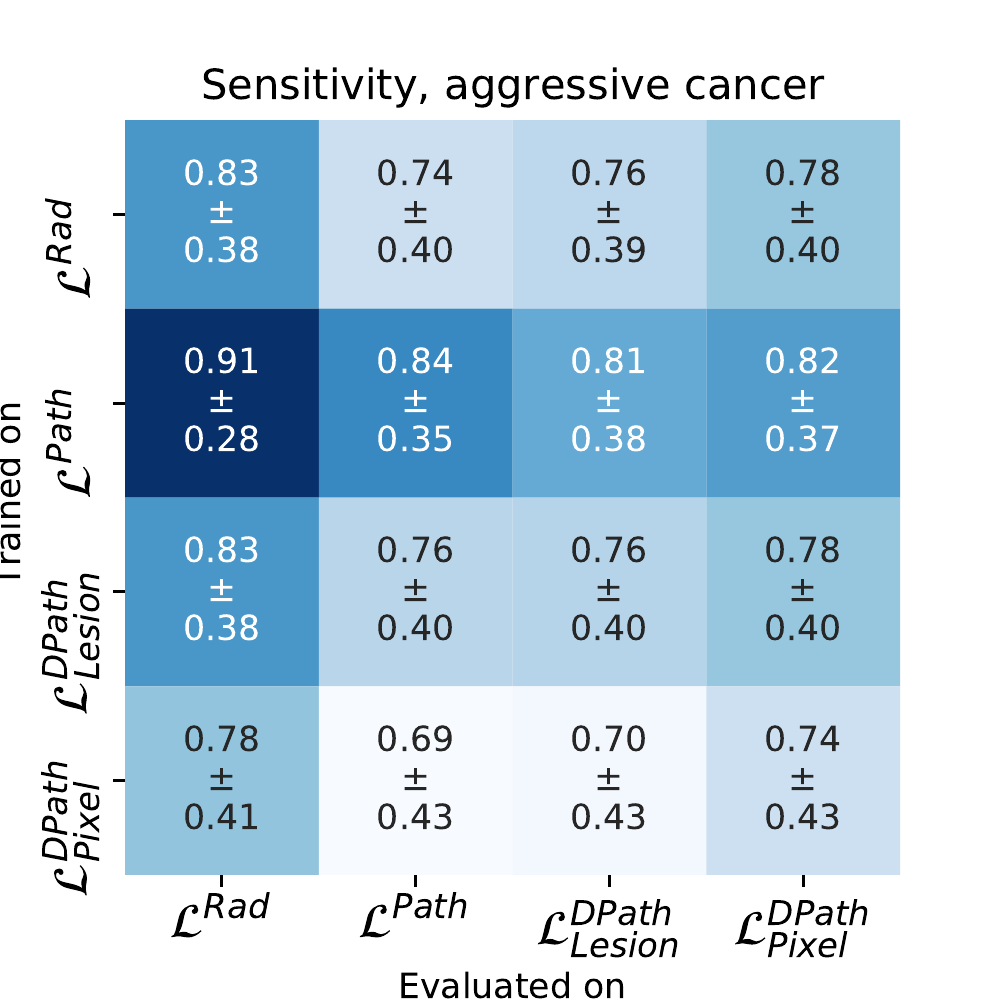}
\includegraphics[trim=0 0 5 31,clip,width = 0.24\textwidth]{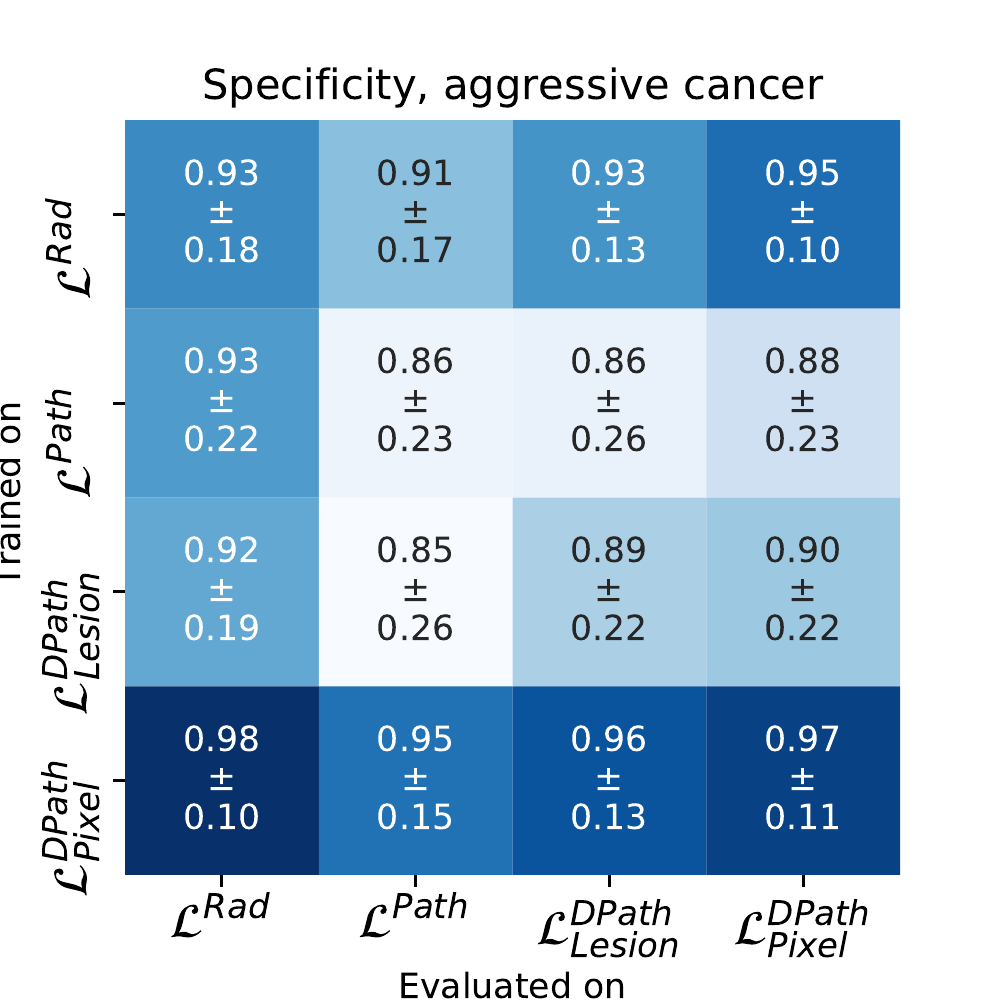}

\begin{subfigure}[b]{.24\linewidth}
  \centering
   \caption{Dice}
  \end{subfigure}
  \begin{subfigure}[b]{.24\linewidth}
  \centering
   \caption{Lesion ROC-AUC}
  \end{subfigure}
\begin{subfigure}[b]{.24\linewidth}
  \centering
   \caption{Lesion Sensitivity}
  \end{subfigure}
\begin{subfigure}[b]{.24\linewidth}
  \centering
  \caption{Lesion Specificity}
  \end{subfigure}
  
\caption{Quantitative comparison between digital radiologist (SPCNet) predictions when trained and evaluated using different label types in cohort C1-test. The top row shows results for cancer detection, while the bottom row shows results for aggressive cancer detection. Darker blue boxes in the 4$\times$4 matrices represent higher evaluation metrics.}
     \label{fig:spcnet_c1_comp}
\end{figure*}

\textbf{Cohort C1-test:} Quantitatively comparing the lesion-level performance of the digital radiologists trained with the different label types in cohort C1-test showed that the type of label used for training has an effect on digital radiologist performance (\figurename~\ref{fig:spcnet_c1_comp}). Digital radiologists trained with radiologist labels ($\mathcal{L}^{Rad}$) had lower Dice overlaps, lower lesion-level ROC-AUCs and lower sensitivities than digital radiologists trained with pathologist labels. Digital radiologists trained with pathologist labels ($\mathcal{L}^{Path}$) had the highest Dice overlaps and sensitivities among all models.

Digital radiologists trained with lesion-level digital patholologist labels ($\mathcal{L}^{DPath}_{Lesion}$) had higher lesion-level ROC-AUCs and sensitivities than
radiologist label-trained models. Oftentimes, $\mathcal{L}^{DPath}_{Lesion}$-trained digital radiologists outperformed pathologist label-trained digital radiologists in lesion-level ROC-AUCs. Digital radiologists trained with pixel-level digital pathologist labels ($\mathcal{L}^{DPath}_{Pixel}$) had higher lesion-level ROC-AUCs than radiologist label-trained digital radiologists and sometimes higher lesion-level ROC-AUCs than pathologist label-trained digital radiologists as well. They also had the highest specificities among all the digital radiologists. 

For all digital radiologists, highest Dice overlaps were achieved when evaluated using radiologist labels ($\mathcal{L}^{Rad}$). This can be attributed to the fact that these cancers captured by $\mathcal{L}^{Rad}$ are more prominent on MRI, making them easier to be learned by the digital radiologists.

\textbf{Cohort C2:} In cohort C2, the digital radiologist trained with radiologist labels ($\mathcal{L}^{Rad}$) had the highest lesion-level ROC-AUC and Dice overlaps (Table~\ref{tab:spcent_c2}). Digital radiologists trained with all pathology labels ($\mathcal{L}^{Path}$, $\mathcal{L}^{DPath}_{Lesion}$ and $\mathcal{L}^{DPath}_{Pixel}$) had slightly lower and similar AUC-ROCs and Dice overlaps. The better performance of $\mathcal{L}^{Rad}$-trained digital radiologists in cohort C2 can be attributed to the fact the evaluation is also with respect to $\mathcal{L}^{Rad}$ in this cohort as other labels are not available.
    
Although the $\mathcal{L}^{Rad}$-trained digital radiologists had the highest ROC-AUCs and Dice overlaps, the digital radiologists trained with lesion-level digital pathologist labels ($\mathcal{L}^{DPath}_{Lesion}$) had the highest sensitivities and the digital radiologists trained with pixel-level digital pathologist labels ($\mathcal{L}^{DPath}_{Pixel}$) had the highest specificities. Moreover, digital radiologists trained with $\mathcal{L}^{DPath}_{Pixel}$ are the only ones that could detect indolent cancer lesions. This can be attributed to the fact, that during training, only digital radiologists trained with $\mathcal{L}^{DPath}_{Pixel}$ get sufficient number of indolent cancer examples.

\begin{table}[!ht]
\centering
\caption{Lesion-level evaluation in cohort C2  of the SPCNet models trained using cohort C1-train. Cohort C2 only had biopsy-confirmed radiologist labels ($\mathcal{L}^{Rad}$), thus all evaluations were with respect to $\mathcal{L}^{Rad}$.}

\begin{tabular}{|C{3cm} c c c c|}
\hline
\hline
\multicolumn{5}{|c|}{Cancer vs.~all (N = 160, number of lesions = 193)} \\
\hline
\hline
Trained with Label Type &  AUC-ROC  &  Dice   &   Sens.  &  Spec. \\
\hline 

$\mathcal{L}^{Rad}$    & \textbf{0.84$\pm$0.29}
   & \textbf{0.39$\pm$0.28}
   
   & 0.70$\pm$0.42
   & 0.85$\pm$0.28 \\ \hline
$\mathcal{L}^{Path}$  & 0.81$\pm$0.33
   & 0.37$\pm$0.27
  
   & 0.70$\pm$0.43
   & 0.73$\pm$0.36\\ \hline
$\mathcal{L}^{DPath}_{Lesion}$    & 0.81$\pm$0.32
   & 0.37$\pm$0.27
  
   & \textbf{0.71$\pm$0.42}
   & 0.78$\pm$0.34\\ \hline
$\mathcal{L}^{DPath}_{Pixel}$   & 0.81$\pm$0.31
   & 0.35$\pm$0.29
 
   & 0.64$\pm$0.45
   & \textbf{0.87$\pm$0.26} \\
\hline
\hline
\multicolumn{5}{|c|}{Aggressive Cancer vs.~all (N = 160, number of lesions = 132)} \\
\hline
\hline
Trained with Label Type &  AUC-ROC  &  Dice   & Sens.  &  Spec. \\ \hline
$\mathcal{L}^{Rad}$ & \textbf{0.89$\pm$0.24}
   & \textbf{0.43$\pm$0.26}
  
   & 0.77$\pm$0.39
   & 0.84$\pm$0.28  \\ \hline
$\mathcal{L}^{Path}$    & 0.87$\pm$0.27
   & 0.41$\pm$0.25
  
   & 0.79$\pm$0.39
   & 0.72$\pm$0.37\\ \hline
$\mathcal{L}^{DPath}_{Lesion}$      & 
    0.87$\pm$0.26
   & 0.42$\pm$0.25
  
   & \textbf{0.81$\pm$0.37}
   & 0.77$\pm$0.36\\ \hline

$\mathcal{L}^{DPath}_{Pixel}$  & 0.88$\pm$0.27
   & 0.40$\pm$0.28
   
   & 0.73$\pm$0.42
   & \textbf{0.85$\pm$0.29}\\
\hline
\hline
\multicolumn{5}{|c|}{Indolent Cancer vs.~all (N = 160, number of lesions = 61)} \\
\hline
\hline
Trained with Label Type &  AUC-ROC  &  Dice   & Sens.  &  Spec. \\ \hline
$\mathcal{L}^{Rad}$ & 0.46$\pm$0.42
  & 0.00$\pm$0.01  
   & 0.02$\pm$0.13
  
   & 0.99$\pm$0.01
 \\ \hline
$\mathcal{L}^{Path}$    & 0.43$\pm$0.43
   & 0.00$\pm$0.00
  
   & 0.00$\pm$0.00
   & \textbf{1.00$\pm$0.00}\\ \hline
$\mathcal{L}^{DPath}_{Lesion}$      & 
    0.43$\pm$0.40
   & 0.00$\pm$0.00
  
   & 0.00$\pm$0.00
   & \textbf{1.00$\pm$0.00}\\ \hline

$\mathcal{L}^{DPath}_{Pixel}$  & \textbf{0.64$\pm$0.40}
   & \textbf{0.12$\pm$0.17}
   & \textbf{0.33$\pm$0.45}
   & 0.94$\pm$0.14\\
\hline
\hline
\end{tabular}
\label{tab:spcent_c2}
\vspace{0.2cm}
\end{table}


\section{Discussion}

In this study, we performed a detailed analysis to (a) compare different prostate cancer labeling strategies, and (b) study the effects these labeling strategies have on the deep learning models (which we refer to as digital radiologists) that are trained with them. Our qualitative and quantitative evaluations indicate that radiologist labels ($\mathcal{L}^{Rad}$) have lower lesion-detection rates than pathology labels (labels on whole-mount histopathology images mapped onto MRI through MRI-histopathology registration), and do not capture the true extent of cancer, in line with prior studies \citep{ahmed2017diagnostic, johnson2019detection, priester2017magnetic}. Subsequently, digital radiologist models trained with $\mathcal{L}^{Rad}$ also have inferior performance when compared to models trained with pathology labels ($\mathcal{L}^{Path}$, $\mathcal{L}^{DPath}_{Lesion}$, $\mathcal{L}^{DPath}_{Pixel}$). Digital pathologist (deep learning method for labeling of Gleason patterns on histopathology images \citep{ryu2019automated}) labels ($\mathcal{L}^{DPath}_{Lesion}$, $\mathcal{L}^{DPath}_{Pixel}$) have high concordance with pathologist labels ($\mathcal{L}^{Path}$). Digital radiologists trained with digital pathologist labels perform with comparable or better accuracy than digital radiologists trained with radiologist or pathologist labels. Moreover, digital radiologists trained with pixel-level digital pathologist labels ($\mathcal{L}^{DPath}_{Pixel}$) can enable selective identification of aggressive and indolent cancer components in mixed lesions, which is not possible by radiologists. Evaluation in both cohorts indicate that the digital radiologists trained with digital pathologist labels have generalizable performance in biopsy as well as radical prostatectomy patients. The trend of digital pathologist label-trained digital radiologists performing better or comparable to human label-trained digital radiologists is irrespective of the model architecture (Table \ref{tab:model_comp_c2}). Thus, digital pathologist labels provide a consistent, standardized, accurate, labor and time-efficient method for training reliable digital radiologists for selective identification of aggressive and indolent prostate cancer.

Digital pathologist labels not only train the most accurate digital radiologists, but using digital pathologist labels to build digital radiologists also helps overcome the challenges associated with generating human-annotated pixel-level histologic grade labels. It is impractical for genitourinary pathologists to manually annotate all prostate pixels with Gleason patterns for a sufficiently large population of patients to train machine learning models. Automated Gleason grading on histopathology images by digital pathologists (a) have excellent performance \citep{ryu2019automated, bulten2020automated}, and (b) have shown to significantly improve Gleason grading by human pathologists \citep{bulten2020artificial}. Digital pathologist labels also improve uniformity in grading by reducing inter- and intra-pathologist variation in Gleason Grade group assignment.

Prior studies \citep{saha2021end, schelb2019classification, sumathipala2018prostate, seetharaman2021automated, cao2019joint, sanyal2020automated, hosseinzadeh2019effect, hosseinzadeh2021deep, yu2020deep, mcgarry2019gleason, de2020deep, bhattacharya2021selective} on developing machine learning methods for prostate cancer detection have used different kinds of labels to develop their models. This is the first study to systematically compare and analyze the effect of different labeling strategies on the performance of automated algorithms for prostate cancer detection on MRI (digital radiologists). We trained four different model architectures (U-Net, branched U-Net, SPCNet and the DeepLabv3+) used in prior studies and tested in two independent cohorts to further emphasize that the effect of the labeling strategies is independent of the model type and the dataset used for testing. Our study showed that the SPCNet architecture outperformed the other architectures, irrespective of the label type used for training. 

Our study has five noteworthy limitations. First, unlike prior studies \citep{hosseinzadeh2021deep}, the number of patients in cohort C1 is relatively small (N=115), primarily due to its uniqueness including registered MRI and histopathology images of radical prostatectomy patients, pixel-level radiologist and pathologist labels, as well as pixel-level digital pathologist labels. Despite its small size, the generalizable performance of the deep learning models on the independent cohort C2 indicate the utility of the dataset. Second, all patients in this study are from a single institution (Stanford University) and single manufacturer (GE Healthcare). Third, our study includes retrospective data and has not been used in prospective evaluation. Fourth, the digital pathologist was trained on prostate biopsy histopathology samples \citep{ryu2019automated}, but was used to generate pixel-level histologic grade labels on whole-mount histopathology images. Despite being trained on biopsy histopathology images, the digital pathologist showed high agreement with the human pathologist on the whole-mount images. Finally, registration errors (\textasciitilde 2~mm on the prostate border and 3~mm inside the prostate) in the MRI-histopathology registration platform \citep{rusu2020registration} may affect small lesions. Excluding lesions of volumes 250~mm$^3$ (6~mm $\times$ 6~mm $\times$ 6~mm) helps focus on aggressive cancer, as small lesions are not deemed to be clinically significant \cite{matoso2019defining, turkbey2019prostate} while helping counter the MRI-histopathology registration errors in cohort C1.  

Identifying and treating aggressive cancer, and reducing over-treatment of indolent cancer are the primary goals of prostate cancer care. A digital radiologist can help standardize radiologist interpretations, and assist clinicians in reliably detecting and localizing aggressive and indolent cancer on prostate MRI. In order to develop a reliable digital radiologist, it is imperative to train it with the best possible labels. Our experiments show that digital pathologist labels are the best way to train digital radiologists not only because they help develop the most accurate digital radiologist models, but also because they circumvent the challenges associated with acquiring pixel-level human-annotated histologic grade labels. A reliable digital radiologist can help prostate cancer care by (1) standardizing radiologist interpretations, (2) helping detect and target aggressive cancers that are currently missed, (3) helping reduce unnecessary invasive biopsies in men without cancer or with indolent cancer, and (4) helping reduce the number of biopsies to detect aggressive cancers by localizing the aggressive cancer components in mixed lesions. 

\section{Conclusion}

Digital pathologist labels generated by deep learning algorithms on prostate histopathology images can help bridge the gap between prostate radiology and pathology by enabling the training of reliable machine learning models, referred to here as digital radiologists, for selective identification of aggressive and indolent prostate cancer on MRI. Digital pathologists have similar performance to pathologists in selective identification of aggressive and indolent prostate cancer on prostate histopathology images. 
Digital pathologist-trained digital radiologists (1) enable selective identification of aggressive and indolent cancer on prostate MRI on a lesion-level as well as on a pixel-level (which is not possible with any human-annotated label type), (2) perform better than radiologist-trained models, (3) perform equally well or better than pathologist-trained models, and (3) circumvent the labor, time, and variability challenges associated with human annotations for training digital radiologist models.

\section*{Acknowledgements}
\addcontentsline{toc}{section}{\numberline{}Acknowledgements}
We acknowledge the following funding sources: Departments of Radiology and Urology, Stanford University, GE Healthcare Blue Sky Award, National Institutes of Health, National Cancer Institute (U01CA196387, to JDB), and the generous philanthropic support of our patients (GS).

\noindent \textbf{Conflict of Interest:} Mirabela Rusu has research grants from GE Healthcare and Philips Healthcare. 


\section*{References}
\addcontentsline{toc}{section}{\numberline{}References}
\vspace*{-20mm}





\bibliography{./refs}      



\bibliographystyle{./medphy.bst}    

\newpage
\huge{Supplementary Material}\\

\large{\textbf{I. MRI parameter acquisition characteristics}} \\
\vspace{3mm}
\begin{table*}[!htb]
\centering
\caption{Description of MRI parameter acquisition characteristics in our two cohorts.}
\begin{tabular}{|p{3.3in}|p{1.05in}|p{1.05in}|}
\hline
 MRI Statistics & Cohort C1 & Cohort C2 \\ \hline
T2w & &  \\ \hline
\hspace{.5cm} Repetition Time (TR) (s) & 3.9-6.3 & 2.0-7.4\\ \hline
\hspace{.5cm} Echo Time (TE) (ms) & 122-130  & 92-150 \\ \hline
\hspace{.5cm} Pixel Size (mm) & 0.27-0.94  & 0.39-0.47 \\ \hline
\hspace{.5cm} Distance between Slices (mm) & 3.00-4.20  & 3.00-4.20\\ \hline
\hspace{.5cm} No. of Slices &   24-43  & 20-43\\ \hline
ADC & & \\ \hline
\hspace{0.5cm} b-values ($s/mm^2$) & [0, 50, 800, 1000, 1200] & [0, 25, 50, 800, 1200, 1400]\\ \hline
\hspace{.5cm} Pixel Size (mm) & 0.78-1.50  & 0.78-1.01\\ \hline
\hspace{.5cm} Distance between Slices (mm) & 3.00-5.20 & 3.00-4.60\\ \hline
\hspace{.5cm} No. of Slices &  15-40 & 14-42\\ \hline

\end{tabular}
\label{tab:dataset}
\end{table*}

\large{\textbf{II. Data preprocessing}}

\textbf{II.A. Registration} \label{sec:registration}

For cohort C1, pre-operative MRI and post-operative histopathology images were registered using the RAPSODI registration platform \citep{rusu2020registration}. This MRI-histopathology registration allows mapping the extent of cancer from histopathology images onto MRI using affine and deformable transformations on corresponding MRI and histopathology images. In addition, for cohort C1, T2w and ADC images were manually registered using affine transformations.

\textbf{II.B. Resampling} \label{sec:resampling}

The T2w and ADC images of all subjects from both cohorts were cropped around the prostate and resampled to have the same pixel-size (0.29mm x 0.29mm) and the same X-Y dimensions (224x224), similar to our prior studies \citep{seetharaman2021automated, bhattacharya2020corrsignet, bhattacharya2021selective}.

\textbf{II.C. MRI Intensity Standardization and Intensity Normalization}
T2w and ADC image-intensities were standardized using a histogram alignment approach \cite{nyul2000new} using average histograms derived from the training set of each MRI sequence independently. Standardized MRI intensities were then z-score normalized, similar to our prior studies \cite{seetharaman2021automated, bhattacharya2020corrsignet}.

\large\textbf{III. Model Architectures}

\textbf{SPCNet:} SPCNet \cite{seetharaman2021automated} is an architecture based on the hierarchical Holistically-Nested Edge Detector (HED) model  \cite{xie2015holistically} that was designed to leverage multiple scales of input features for edge detection. SPCNet has 2 separate encoders for T2w and ADC images respectively, with each encoder taking in three adjacent MRI slices. The outputs from each encoder are concatenated and go through more convolutional layers. Then, the outputs of those convolutional layers are fused with side outputs from both encoders as well as from the post-concatenation convolutional layers. This fused final output is used as input to the final softmax layer that predicts the probability of each class for each pixel. 

\noindent\textbf{U-Net:} U-Net \cite{ronneberger2015u} is a commonly used deep learning model for biomedical image segmentation tasks including prostate cancer detection \cite{schelb2019classification, sanyal2020automated}. The network architecture of U-Net consists of a traditional ``contracting'' path of convolution layers, or encoder, followed by an``expanding'' mirror set of convolutional layers, known as the decoder, that outputs the segmentation map. In addition to the main path, ``skip-connections'' between corresponding encoder and decoder layers allow the decoder to utilize additional features of the input directly from the encoder. Three adjacent slices of T2w images and three slices of the corresponding ADC images were input into the U-Net model as image channels with 6 input channels in total. 

\noindent\textbf{Branched U-Net (BrU-Net):} 
A variant of the vanilla U-Net architecture, which we call the branched U-Net (BrU-Net), was used in our experiments. The BrU-Net incorporates the changes that SPCNet incorporates to the baseline HED architecture, i.e., BrU-Net has two separate encoders for the T2w and ADC images, with each encoder taking in three adjacent MRI slices. Decoder has identical layers to that of the original U-Net but has skip-connection inputs from both branches.

\noindent\textbf{DeepLabv3+:} DeepLabv3+ \cite{chen2018encoder} is a deep learning model for semantic segmentation that builds on prior DeepLab architectures by including atruous convolutions, spatial pyramid pooling, and integrating a decoder that is better at segmentating boundary details. The DeepLabv3+ architecture formed the backbone of the FocalNet model for prostate cancer detection and Gleason grade prediction \citep{cao2019joint}. For our experiments, the encoder of DeepLabv3+ takes as input one slice of T2w and one slice of ADC per example.

\large\textbf{IV. Evaluation Metrics}\\

The following metrics were used for analysis:
$$Dice =  \frac{2 * TP}{2 * TP + FP + FN}$$ 
$$Sensitivity =  \frac{TP}{TP + FP}$$ 
$$Specificity =  \frac{TN}{TN + FP}$$ 

where TP are the true positive and FP are the false positive predictions. The Dice coefficient was computed on a pixel-level, wheras the sensitivities and specificites were computed on a lesion-level using the predicted and ground truth labels. In addition, predicted probabilities were used to compute the lesion-level area under the receiver operating characteristics (ROC-AUC) curves.  

\end{document}